\documentclass{article}
\title{Concurrent Realizability on Conjunctive Structures}
\author{Emmanuel Beffara \and F\'elix Castro \and Mauricio Guillermo}
\usepackage[a4paper,width=18cm,height=23cm]{geometry}
\usepackage{macros}
\begin{document}
\maketitle

\section{Introduction}

The point of this work is to explore axiomatisations of concurrent computation
using the technology of proof theory and realizability.

\emph{Make sentences and paragraphs \dots}

Justifications for this work:
\begin{itemize}
  \item Type systems for concurrent calculi are diverse and heterogeneous,
    there is no consensual common logical foundation.
  \item The situation is similar regarding semantics of processes.
\end{itemize}
Arguments for this approach:
\begin{itemize}
  \item Realizability is a fruitful and flexible tool for relating
    intuitionistic and classical logics with computation models, including
    control, effects, etc.
  \item It does not assume a pre-existing logical structure but only a basic
    notion of composition.
  \item The framework of implicative structures and related systems unifies
    computational objects (programs) and types (behaviours) in a common
    structure.
  \item This is especially relevant when considering concurrent processes
    where it is customary to use process both as specifications and
    implementations.
\end{itemize}

\section{Definitions}

\subsection{Processes and processes with fusions}

We start by describing the language of processes which are the realizers of the concurrent realizability. We define the \emph{Processes With Fusions} (PWF) which are pairs $(P, e)$ where $P$ is a $\pi$-term and $e$ is a \emph{fusion} identifying channels. Those fusions are formalized as equivalence relations on $\NN$ (i.e.: on the set of names) where each equivalence class is finite. 

Firstly we remember the definitions of $\pi$-calculus without additions and replication. We restrict to this calculus and the corresponding fragment of linear logic following the approach of Honda \& Yoshida on his article \emph{Combinatory  Representation  of  Mobile  Processes}  \cite{Yoshida-Honda-combinators}. For the actions of emissions (resp. receptions) we choose to take actions which are binders for the names which are emitted (resp. received), following the symmetric approach given by Beffara on his PhD thesis \cite{Beffara-thesis} . 

After that we define the (PWF)-processes and redefine the usual constructions of $\pi$-calculus for the processes with fusions.

%On the annex section we discuss different definitions of Labelled Transition Systems (LTS) for the (PWF). However, none of the results we exhibit on the sections 1, 2 and 3 depends upon the definition of a (LTS) on (PWF)-processes.  

\begin{definition}\label{df:Processes-and-equiv}
  The set $\Pi$ of $\pi$-processes is defined by the following grammar:
  \[
    \begin{array}{rcc|c|c|c}
      P, Q&::=&\one&(P|Q)&u^{\varepsilon}(\vec{x}).P&(\nu x)P%!P
    \end{array}
  \]
  where $\vec{x}$ is a vector of names, $x$ and $u$ are names and $\varepsilon$ is a polarity. The expressions of the shape $u^\varepsilon(\vec{x})$ are called \emph{actions} and whenever it is not necessary to specify $u, \varepsilon$ and $\vec{x}$, we write simply $\alpha, \beta, \gamma, \dots$ to denote action. 

  The \emph{free names} of a process $P$ is defined by induction:
  \begin{itemize}
    \item $\Fn(\one):=\emptyset$.
    \item $\Fn(P|Q):=\Fn(P)\cup \Fn(Q)$.
    \item $\Fn(u^\varepsilon(\vec{x}).P):=(\Fn(P)\setminus\{\vec{x}\})\cup \{u\}$.
    \item $\Fn((\nu x)P):=\Fn(P)\setminus\{x\}$. 
  \end{itemize}
  As usual, processes are considered up to $\alpha$-conversion. 
  The \emph{structural equivalence} on processes is defined by the following rules:
  \begin{itemize}
    \item $(\Pi, |, \equiv)$ is an abelian monoid, i.e.: $\one|P\equiv P$, $P|Q\equiv Q|P$ and $(P|Q)|R\equiv P|(Q|R)$.
    \item $P\equiv (\nu x)P$ and $P|(\nu x)Q\equiv (\nu x)(P|Q)$ whenever $x\notin \Fn(P)$. Moreover, $(\nu x)(\nu y)P\equiv (\nu y)(\nu x)P$ and we write $(\nu xy)P:=(\nu x)(\nu y)P$.
    \item $\equiv$ is a congruence, i.e.:
      \begin{center}
        \begin{prooftree}
          P\equiv P'
          \justifies
          u^\varepsilon(\vec{x}).P\equiv u^\varepsilon(\vec{x}).P'
        \end{prooftree}
        \hfil
        \begin{prooftree}
          P\equiv P'
          \justifies
          P|Q\equiv P'|Q
        \end{prooftree}
        \hfil
        \begin{prooftree}
          P\equiv P'
          \justifies
          (\nu x)P\equiv (\nu x)P'
        \end{prooftree}
      \end{center}
  \end{itemize}
\end{definition}

\begin{definition}\label{df:substitutions}
  A substitution is a function $\sigma\in\NN^\NN$. Given a substitution $\sigma$ and a set $X\subseteq \NN$ we denote by $\sigma\setminus X$ the restriction $\sigma_{\upharpoonright X^c}:X^c\to\NN$. 
  We define the action of substitutions over $\pi$-processes by means~of:
  \begin{itemize}
    \item $\one_\sigma:= \one$.
    \item $(P|Q)_\sigma:=P_\sigma|Q_\sigma$.
    \item $(u^\varepsilon(\vec{x}).P)_\sigma:=u_\sigma^{\varepsilon}(\vec{x}).P_{\sigma{\setminus}\{\vec{x}\}}$ possibly renaming the bound names $\vec{x}$ to prevent the capture of free names.
    \item $((\nu x).P)_\sigma:=(\nu x).P_{\sigma{\setminus}\{x\}}$ possibly renaming the bound name $x$ to prevent the capture of free names.
      %\item $(!\alpha.P)_\sigma:=! (\alpha.P)_\sigma$. 
  \end{itemize}
\end{definition}
Intuitively, fusions are connections between channels. The fundamental properties of connections are that every channel is \emph{de facto} connected to itself and that if $a$ is connected to $b$ and $b$ is connected to $c$, then $a$ and $c$ are connected. Unlikely what happens on the Beffara's \emph{polarized} $\pi$ \emph{calculus}~\cite{Beffara-thesis}, we present connections that are not oriented. Then, the connection relationship is a symmetric relation. We conclude that the connection relationship are \emph{equivalence relations} on names.  

For technical reasons we ask that each name be fused with a finite number of names (i.e.: the equivalence classes are finite sets). On the other hand, we allow that infinitely many names are fused to another name (i.e.: there are infinitely many non trivial equivalence classes). 

We observe that on the Beffara's polarized $\pi$-calculus \cite{Beffara-thesis}, to give a process the type $\vdash A^\perp, A$, it is needed to assign to $A$ and $A^\perp$ an interface of polarized names,  which determines the parallel composition of finite fusions that have type $\vdash A^\perp, A$. 
Because our fusions potentially involve infinitely many channels,  we can give to a unique process (in fact a fusion) the type $\vdash A^\perp, A$. 
\begin{definition}\label{df:operations-with-fusions} 
  The set of fusions is $\mathcal{E}:=\{e\subseteq \NN^2\ |\ e \text{ is an equivalence relation and }(\forall x{\in}\NN)\ [x]_e \text{ is finite}\}$. The minimum fusion is $\Delta_\NN:=\{(x,x)\ |\ x\in\NN\}$. Given $e, f\in\mathcal{E}, x\in\NN$ and $\sigma\in\NN^\NN$ we define also:
  \begin{itemize}
    \item $[x]_e:=\{y\in \NN\ |\ x\simf{e} y\}$ (the equivalence class of $x$). Also define \emph{the domain of} $e$ as $|e|:=\{x\in\NN\ |\ [x]_e\neq \{x\}\}$.
    \item $e\vee f = ef = \operatorname{tcl}(e\cup f)$ where $\operatorname{tcl}$ is the transitive closure (since $e\cup f$ is simmetric and reflexive, $e\vee f\in\mathcal{E}$). 
    \item Given $X\subseteq \NN$, we define $e\cap X:=(e\cap X^2)\cup \Delta_\NN$ and $e\setminus X:=e\cap X^c = (e\setminus (X^c)^2)\cup\Delta_\NN$.
  \end{itemize}
\end{definition}

\begin{proposition}\label{pr:fusions-as-lattice}\  
  \begin{enumerate}
    \item Given fusions $e, f$ and sets $X, Y$, $e\vee f= e f$, $e\cap X$ and $e\setminus X$ are fusions.
    \item $(e\cap X)\cap Y= e\cap (X\cup Y)$ and $(e\setminus X)\setminus Y=e\setminus (X\cup Y)$.
    \item The set of fusions together with $\vee, \cap$ is a lattice $(\mathcal{E}, \vee, \cap)$ with the following properties:
      \begin{enumerate}
        \item $\Delta_\NN$ is the minimum element of $\mathcal{E}$. However, $\mathcal{E}$ has no maximum element. 
        \item Given $e, f, g\in\mathcal{E}$, $(e\cap g)\vee (f\cap g)\subseteq (e\vee f)\cap g$ and in general the converse inclusion is false. 
        \item Given $e, f, g\in\mathcal{E}$, $(e\cap f)\vee g\subseteq (e\vee g)\cap (f\vee g)$ and in general the converse inclusion is false 
      \end{enumerate}
  \end{enumerate}   
\end{proposition}
\begin{proof} 
  \begin{enumerate}
    \item[1,2] are straightforward.
    \item[3] The first sentence is a direct verification. For the second observe that since $(e\cap g)\cup (f\cap g)=(e\cup f)\cap g\subseteq (e\vee f)\cap g$, then $(e\cap g)\vee (f\cap g)\subseteq (e\vee f)\cap g$. The converse inclusion is in general false: consider the counterexample $e:=\{x_0\leftrightarrow x_1\}, f:=\{x_1\leftrightarrow x_2\}$ and $g:=\{x_0\leftrightarrow x_2\}$. Clairly $x_1\overset{(e\vee f)\cap g}{\longleftrightarrow} x_2$ but since $e\cap g=f\cap g=\Delta_\NN$ then $(e\cap g)\vee (f\cap g)=\Delta_\NN$ and hence $(x_0, x_2)$ are not related by $(e\cap g)\vee (f\cap g)$. For the third sentence the reasoning is similar, with the same relations $e, f, g$ and the couple $(x_1, x_2)$ as a counterexample for the converse inclusion.    
  \end{enumerate}   
\end{proof}

\begin{definition}\label{df:operations-with-substitutions}\  
  \begin{itemize}
    \item Given $e\in\mathcal{E}$ and $x\in\NN$ we define $x^\bullet_e:=\min[x]_e$ and $x^*_e:=\begin{cases}
        x & \text{ if } [x]_e= \{x\}\\
        \min([x]_e\setminus\{x\}) & \text{ if } [x]_e\neq \{x\}\\
      \end{cases}$ 

      Define $\sigma_e:\NN\to\NN$ s.t. $\forall x{\in}\NN\quad \sigma_e(x):=x_e^\bullet$. 
    \item Given a substitution $\sigma\in\NN^\NN$ and a fusion $e\in\mathcal{E}$, we define $e_\sigma:=\operatorname{trcl}\big(\{(\sigma(x), \sigma(y))\ |\ x\simf{e} y\}\big)$. 

      %Given $x\in\NN$, we define $\sigma_{(e,x)}\in \NN^\NN$ s.t. $\sigma_{(e,x)}(z):=\left.\begin{cases}
      %  \sigma(z) & \text{ if } z\neq x\\
      %  x^*_e & \text{ if } z=x\\
      %\end{cases}\right\}=: \sigma\{x:=x^*_e\}$. In particular, $\id_{(e,x)}=\{x:=x^*_e\}$.  
  \end{itemize}
\end{definition}

\begin{lemma}\label{lm:sigma-e/sigma-e/x}
  Let us consider $e\in\mathcal{E}$ and $x\in\NN$. Then $\sigma_{e\setminus\{x\}}\circ \{x:=x^*_e\}=\{x:=x^*_e\}\circ \sigma_e$
\end{lemma}
\begin{proof}
  Observe that
  \[
    \sigma_{e{\setminus}\{x\}}(z)=\begin{cases}
      z^\bullet_e& \text{ if }z\not\simf{e}x\\
      x^*_e& \text{ if }z\neq x\simf{e} z\\
      x & \text{ if } z=x\\
    \end{cases}
  \] 
  Let us denote by $\tau:=\sigma_{e\setminus\{x\}}\circ  \{x:=x^*_e\}$ and $\rho:=\{x:=x^*_e\}\circ \sigma_e$. To prove that $\tau=\rho$ we have two cases:
  \begin{itemize}
    \item If $z\not\simf{e} x$, then $\tau(z)=\sigma_{e{\setminus}\{x\}}(z)=z^\bullet_e$ and $\rho(z)=\{x:=x^*_e\}(z^\bullet_e)=z^\bullet_e$, thus proving the result.
    \item If $z\simf{e}x$ then $\tau(z)=x^*_e$, whether $z=x$ or not. On the other hand, $\rho(z)=\{x:=x^*_e\}(z^\bullet_e)=\{x:=x^*_e\}(x^\bullet_e)$ and we have two cases of analysis:
      \begin{itemize}
        \item If $x\neq x^\bullet_e$, then $x_e^\bullet=x_e^*$ (because $x^\bullet_e=x^*_e$ iff $x\neq x^\bullet_e$) and hence $\{x:=x^*_e\}(x^\bullet_e)=x^\bullet_e=x^*_e$.
        \item If $x=x^\bullet_e$, then $\{x:=x^*_e\}(x^\bullet_e)=x^*_e$. 
      \end{itemize}
  \end{itemize}
\end{proof}

\begin{definition}\label{df:equivalence-between-substitutions}
  Let us consider $\sigma, \tau:\NN\to\NN$. We say that $\sigma\sim\tau$ (we say they are equivalent) iff exists a finite permutation $\rho$ s.t. $\sigma =\rho^{-1}\circ\tau\circ\rho$. 
\end{definition}

\begin{lemma}\label{lm:order-equivalence-blocking+renaming}
  Let us consider $x, y\in\NN$, $e\in\mathcal{E}$ and $\sigma:=\{y:=y^*_e\}, \tau:=\{x:=x^*_{e\setminus\{y\}}\}, \sigma':=\{x:=x^*_e\}$ and $\tau':= \{y:=y^*_{e\setminus\{x\}}\}$. Then:
  \begin{enumerate}
    \item If $[x]_e=[y]_e=\{x,y\}$, then $\tau\circ\sigma=\{y:=x\}$ and $\tau'\circ\sigma'=\{x:=y\}$.
    \item If $x\not\simf{e} y$ or $[x]_e=[y]_e\subset \{x,y\}$, then $\tau\circ\sigma=\tau'\circ\sigma'$.
  \end{enumerate}
  In both cases we conclude that $\tau\circ\sigma\sim \tau'\circ \sigma'$.  
\end{lemma}
\begin{proof}
  For $x=y$ is clear. If $x\neq y$ there are two cases:
  \begin{enumerate} \item $x\simf{e} y$. This case splits into two subcases:
    \begin{enumerate}
      \item $[x]_e=[y]_e=\{x,y\}$. Observe that $x^*_e=y, y^*_{e\setminus\{x\}}=y, y^*_e=x$ and $x^*_{e\setminus\{y\}}=x$ and then $\tau\circ\sigma=\sigma=\{y:=x\}$ and $\tau'\circ \sigma'=\sigma'=\{x:=y\}$ which are equivalent by means of the permutation (in fact a transposition)~$(x y)$. 
      \item $[x]_e=[y]_e\subset \{x,y\}$. Let us consider $z=\min ([x]_e{\setminus}\{x,y\})$. Without loss of generality we can suppose that $x<y$. Then we have three subcases:
        \begin{enumerate}
          \item $z<x<y$. In this case $x^*_e=z$, $y^*_{e\setminus\{x\}}=z$ and $y^*_{e}=z$, $x^*_{e\setminus\{y\}}=z$. Thus $\sigma=\tau'=\{y:=z\}$, $\sigma'=\tau=\{x:=z\}$ and hence $\tau\circ\sigma = \tau'\circ\sigma'$ (which are equivalent by means of the identity). 
          \item $x<z<y$. In this case $x^*_e=z$, $y^*_{e\setminus\{x\}}=z$ and $y^*_{e}=x$, $x^*_{e\setminus\{y\}}=z$. Thus
            $\sigma=\{y:=x\}, \tau=\{x:=z\}$,\\ $\sigma'=\{x:=z\}$ and $\tau'=\{y:=z\}$ from which we get $\tau\circ\sigma=\{y:=z, x:=z\}=\tau'\circ\sigma'$. 
          \item $x<y<z$. In this case $x^*_e=y$, $y^*_{e\setminus\{x\}}=z$ and $y^*_{e}=x$, $x^*_{e\setminus\{y\}}=z$. Thus $\sigma=\{y:=x\}$,\\ $\tau=\{x:=z\}, \sigma'=\{x:=y\}$ and $\tau'=\{y:=z\}$ from which we get $\tau\circ\sigma =\{y:=z, x:=z\}=\tau'\circ\sigma'$. 
        \end{enumerate}
    \end{enumerate}
  \item Suppose that $x\not\simf{e} y$. Then we have that $x^*_e=x^*_{e\setminus\{y\}}=x'$ and $y^*_e=y^*_{e\setminus\{x\}}=y'$ for some $x'\not\simf{e} y'$. Thus $\sigma=\tau'\{y:=y'\}$, $\sigma'=\tau=\{x:=x'\}$ from which we get $\tau\circ \sigma =\{x:=x', y:=y'\}=\tau'\circ \sigma'$. 
\end{enumerate}
\end{proof}

\begin{definition}\label{df:equivalence-of-PWF}
  The set of processes with fusions (PWF) is $\overline{\Pi}:=\Pi\times \mathcal{E}$. Given $(P, e), (Q, f)\in\overline{\Pi}$ we say that they are \emph{structurally equivalent} iff $P_{\sigma_e}\equiv Q_{\sigma_f}$ and $e=f$ and we write $(P, e)\equiv (Q, f)$.

  We define the free names of a (PWF) $(P,e)$ by means of $\Fn(P,e):=N_P\cup |e|$ where $N_P:=\{x\in\NN\ |\ [x]_e\cap \Fn(P)\neq \emptyset\}$.
\end{definition}

Thus the free names of $(P,e)$ are all the names which are equivalent to some free name of $P$ along with all the names which are on the domain of $e$.  Observe that since $\Fn(P)$ is finite and the equivalence classes of $e$ are finite, then $N_P$ is finite.

\begin{definition}\label{df:PWF-setminus-a-set}
  Let us consider a PWF $(P,e)$ and a set $X\subseteq \NN$. We denote by $(P,e)\setminus X$ the PWF $(P, e\setminus X)$. 
\end{definition}

\begin{definition}\label{df:iota-immersions}
  We fix injections  $\iota_1, \iota_2:\NN\to\NN$ such that $\iota_1(\NN)\cap\iota_2(\NN)=\emptyset$ and $\iota_1(\NN)\cup \iota_2(\NN)=\NN$. For instance, we can take $\iota_1(n):=2n+1$ and $\iota_2(n):=2n$. We use the following notations: $\NN^i:=\iota_i(\NN)$, $n.i:=\iota_i(n)$ and the substitution $P^i:=P_{\iota_i}$ for $i=1,2$.
\end{definition}  

\subsection{The \texorpdfstring{$\nu-$}\ binder}

Following we define a $\nu$-binder for PWF. The result of $(\nu x)(P,e)$ must be $(Q,f)$ where $x\notin \Fn(Q,f):=N_Q\cup |f|$. Let us consider $(\nu x)(x^\uparrow(), x\leftrightarrow y)$. Even if $x$ must be bind on $(\nu x)(x^\uparrow(), x\leftrightarrow y)$, since $x$ is fused with $y$, the reasonable definition for  $(\nu x)(x^\uparrow(), x\leftrightarrow y)$ is $(y^\uparrow(), \Delta_\NN)$. On the other hand, if we have $(\nu x)(x^\uparrow(), \Delta_\NN)$, what we expect is simply bind $x$ on the process side, i.e.:  $(\nu x)(x^\uparrow (),  \Delta_\NN)=((\nu x)x^\uparrow(), \Delta_\NN)$. 

Both situations are covered by the following definition by cases:
\[
  (\nu x)(P, e):=\begin{cases}
    (P\{x:=x^*_e\}, e\setminus\{x\})& \text{ if } x\in |e|\\
    ((\nu x)P, e) & \text{ otherwise }\\
  \end{cases} 
\]
The following definition includes up to $\alpha$-equivalence both cases into a single sentence:

\begin{definition}\label{df:substitution-block-names}
  %Let us define $\varphi:\mathcal{E}\times\NN\to \NN^\NN\times \mathcal{E}$ s.t. $\varphi(e,x):=(\id_{(e,x)}, e\setminus\{x\})= (\{x:=x^*_e\}, e\setminus\{x\})$.
  Let us consider $(P, e)\in\overline\Pi$ and $x\in\NN$. We define $(\nu x)(P, e):=((\nu x)P\{x:=x^*_e\}, e{\setminus}\{x\})$.  
\end{definition}

\begin{remark}\label{rk:definition-emmanuel}
  Observe that if $[x]_e\neq \{x\}$ then $((\nu x)P\{x:=x^*_e\}, e{\setminus}\{x\})\equiv (P\{x:=x^*_e\}, e{\setminus}\{x\})$. Indeed: since $[x]_e\neq \{x\}$ then $x\neq x^*_e$ which implies $x\notin \Fn(P\{x:=x^*_e\})$ and hence $(\nu x)P\{x:=x^*_e\}\equiv P\{x:=x^*_e\}$.

  On the other hand, if $[x]_e=\{x\}$, then $x=x^*_e$ and then $(\nu x)(P, e)=((\nu x)P, e)$. 
\end{remark}

\begin{corollary}\label{cr:commutation-nu-abstraction}
  Let us consider a (PWF)-process $(P, e)$ and names $x, y$. Then $(\nu x)(\nu y)(P, e)\equiv_\alpha (\nu y)(\nu x)(P, e)$.
  More generally, given a finite set $X=\{x_1, \dots, x_n\}\subseteq \NN$ and any permutation $\rho$ of $n$ elements, then \[(\nu x_1) \dots (\nu x_n)(P, e)\equiv_\alpha (\nu x_{\rho(1)})\dots(\nu x_{\rho(n)})(P, e)\] 
\end{corollary}
\begin{proof}
  Observe that $(\nu x)(\nu y)(P, e)=(\nu x)((\nu y)P\{y:=y^*_e\},e{\setminus}\{y\})=((\nu xy)P\{y:=y^*_e\}\{x:=x^*_{e\setminus\{y\}}\}, e{\setminus} \{x,y\})$. By {\bf Lemma} \ref{lm:order-equivalence-blocking+renaming} this is $\alpha$-equivalent to $((\nu yx)P\{x:=x^*_e\}\{y:=y^*_{e\setminus\{x\}}\}, e{\setminus}\{x,y\}) =$\\ $(\nu y)((\nu x)P\{x:=x^*_e\}, e{\setminus}\{x\})=(\nu y)(\nu x)(P, e)$. The generalization to finite sets is straightforward. 
\end{proof}

\begin{definition}\label{df:general-nu-definition}
  Given a \emph{finite} set of names $X:=\{x_1, \dots, x_k\}$ s.t. $x_1< \dots <x_k$ we define $(\nu X)(P,e):=(\nu x_k)\dots (\nu x_1)(P,e)$. If we consider processes up to $\alpha$-conversion the choice of an order on $X$ is unnecessary.
\end{definition}

The latter gives --at least of $\alpha$-equivalence-- a \emph{definition of the binder} $(\nu X)$ for any \emph{finite set} $X$. 
In order to \emph{define a binder} $(\nu X)$ where $X$ is an \emph{infinite set}, it is necessary at least to bind the names of $X_P:=\Fn(P)\cap X$, which is a finite set. However, it is possible that due to the substitutions defined by the $\nu$-operator, new names appear that must also be bind. For instance, let us consider the following case for $P=n.1^\uparrow()$ and try to intuitively compute  
\begin{center}
  $(\nu_1)(P, n.1\leftrightarrow m.1; k.1\leftrightarrow k.2)=(\nu_1)(n.1^\uparrow(), n.1\leftrightarrow m.1; k.1\leftrightarrow k.2)$  where $\nu_1$  means  $\nu\NN^1$
\end{center}
The set $X_P=\{n.1\}$ and $(\nu n.1)(n.1^\uparrow(), n.1\leftrightarrow m.1; k.1\leftrightarrow k.2)=(m.1^\uparrow(), k.1\leftrightarrow k.2)$. It is clear that a reasonable definition of $(\nu_1)(P, e)$ must give a process whose free names are not members of $\NN^1$. Then we need to bind also $m.1$ on the process side and $k.1$ on the fusion side. 

The following does the job: $(\nu_1)(n.1^\uparrow(), n.1\leftrightarrow m.1; k.1\leftrightarrow k.2):= (\nu n.1, m.1)(n.1^\uparrow(), n.1\leftrightarrow m.1; k.1\leftrightarrow k.2)\setminus \NN^1=(\nu m.1)(\nu n.1)(n.1^\uparrow(), n.1\leftrightarrow m.1; k.1\leftrightarrow k.2)\setminus\NN^1 =  (\nu m.1)(m.1^\uparrow(), k.1\leftrightarrow k.2)\setminus\NN^1 = ((\nu m.1)m.1^\uparrow(), k.1\leftrightarrow k.2)\setminus\NN^1 = ((\nu m.1)m.1^\uparrow(),\Delta_\NN)$. 

It is clear that the fusion side should be uniformly bind over $\NN^1$ at the end of the calculation. But the exact finite set of names to pass to the operator $\nu$ demands to treat the substitution hereditarily: 
\begin{definition}\label{df:hereditary-closure-finite-set}
  Let us consider a set of names $X$ and a finite subset $S$. We recursively define the following sequence $\Big(S_n,\sigma_n\Big)\in\mathcal{P}(X)\times \mathcal{E}$:
  \begin{itemize}
    \item $S_0:=S$, $\sigma_0:=\emptyset$.
    \item Suppose that $S_n=\{s_1, \dots, s_k\}$ (in increasing order). Let us define $\sigma_{n+1}:S_n\to\NN$ given by:
      \[
        \begin{array}{rcl}
          t_1&:=&s^*_{1,e}\\
          \vdots&&\vdots\\
          t_{h+1}&:=&s^*_{h+1,e\setminus\{s_1, \dots, s_h\}}\\
          \vdots&&\vdots \\
          \sigma_{n+1}&:=&\{s_{k}:=t_k\}\circ\dots\circ\{s_1:=t_1\}
        \end{array}
      \]
      And define $S_{n+1}:=S_n\cup (\{t_1, \dots, t_k\}\cap X)$. 
  \end{itemize}
\end{definition}

\begin{proposition}\label{pr:hereditary-closure-finite-set}
  With the hypothesis and the notations of {\bf Definition} \ref{df:hereditary-closure-finite-set} we have:
  \begin{enumerate}
    \item\label{pr:hereditary-closure-finite-set-it1} For all $n\in\NN$ we have $S_n\subseteq [S]_e:=\{y\in \NN\ |\ \exists x{\in} S\ \ x\simf{e} y\}$ and $S_{n}\subseteq S_{n+1}$
    \item\label{pr:hereditary-closure-finite-set-it2} $[S]_e:=\displaystyle\bigcup_{x\in S}[x]_e$ is finite and hence there exists $m\in\NN$ s.t. $\forall n>m\ \ S_n=S_m$. 
    \item\label{pr:hereditary-closure-finite-set-it3} Let us define $m:=\min \{n\in\NN\ |\ S_n=S_{n+1}\}$. Suppose that $S_m:=\{s_1, \dots, s_k\}$. Then, for all $h\in\{1, \dots, k\}$ we have one of the following:
      \begin{enumerate}
        \item\label{pr:hereditary-closure-finite-set-it3-a} $t_h\in X$ and $s_h<t_h$.
        \item\label{pr:hereditary-closure-finite-set-it3-b} $t_h\notin X$ and $\forall j>h$ if $s_j\simf{e} s_h\Longrightarrow t_j=t_h$. 
      \end{enumerate}
    \item\label{pr:hereditary-closure-finite-set-it4} Let us consider $m:=\min\{n\in\NN\ |\ S_n=S_{n+1}\}$ and suppose that $[x]_e\cap X^c\neq\emptyset$ for all $x\in S_m$. Then $\sigma_{m+1}:S_m\to X^c$. In particular, if $X=\NN^1$, then $\sigma_{m+1}:S_m\to\NN^2$. Moreover $\sigma_{m+1}$ is constant over each equivalence class of $e\cap S_m$. 
  \end{enumerate}
\end{proposition}
\begin{proof}\ 
  \begin{enumerate}
    \item By induction, since $S_0\subseteq [S]_e$ and $S_n\subseteq [S]_e\Longrightarrow S_{n+1}\subseteq [S]_e$. Also by definition $S_n\subseteq S_{n+1}$.
    \item Since $e$ is a fusion, each equivalence class is finite and then $[S]_e$ is a finite union of finite sets. Then the increasing sequence $S_n\subseteq [S]_e$ must stabilize from some $n\in\NN$, which ends the proof.
    \item By hypothesis $S_{n+1}=S_n$, then $\{t_1, \dots, t_k\}\cap X\subseteq S_n$. 

      If $t_h\in X$, then $t_h\in S_n$. Since $t_h=s^*_{h, e\setminus\{s_1, \dots, s_{h-1}\}}$, then $t_h\neq s_1, \dots, s_h$. Then we deduce that $t_h=s_j$ with $j>h$. Since $S_n$ is increasingly ordered we deduce that $s_h<t_h$. 

      If $t_h\notin X$ then for all $j>h$ $t_h\in [s_{j}]_{e\setminus\{s_1, \dots, s_j\}}$ and then $t_j=\min [s_{j}]_{e\setminus\{s_1, \dots, s_j\}} = t_h$. 
    \item Let us consider $S_m=C_1\biguplus\dots\biguplus C_l$ the decomposition of $S_m$ on $e\cap S_m$-equivalence classes. Let us consider $C_i=\{s_{i,1}, \dots, s_{i,r_i}\}$ where $s_{i,1}<\dots <s_{i,r_i}$. If we suppose that $t_{i,k}\in X$ for all $k\in\{1, \dots, r_i\}$, by {\bf Proposition} \ref{pr:hereditary-closure-finite-set}.\ref{pr:hereditary-closure-finite-set-it3-a} we deduce $s_{i,r_i}<t_{i,r_i}$ which leads a contradiction because otherwise $t_{i,r_i}\in S_{m+1}\setminus S_m$ and then $S_m\neq S_{m+1}$. Then  there exists a $k_i$ s.t. $t_{k_i}\notin X$ and by {\bf Proposition} \ref{pr:hereditary-closure-finite-set}.\ref{pr:hereditary-closure-finite-set-it3-b} we have that $\forall j>k_i\ \ t_{i,j}=t_{i,k_i}$. By definition, $\sigma_{m+1}$ on $C_i$ is the composition of the following substitutions:
      \[
        \begin{array}{rcl}
          s_{i,1}&\mapsto&s_{i,2}\\
          \vdots&&\vdots\\
          s_{i,k_i-1}&\mapsto&s_{i,k_i}\\
          s_{i,k_i}&\mapsto&t_{i,k}\\
          \vdots&&\vdots\\
          s_{i,r_i}&\mapsto&t_{i,k_i}\\
        \end{array}   
      \]
      which is the constant substitution $\sigma_{m+1}(x)=t_{i,k_i}\in X^c$ for all $x\in C_i$. Since this reasoning does not depends upon $C_i$, we get $\sigma_{m+1}:S_m\to X^c$ and $\sigma_{m+1}$ is constant over each equivalence class $C_i$.
  \end{enumerate}
\end{proof}

\begin{definition}\label{df:hereditary-closure-finite-set-II}
  Let us consider a set of names $X$ and a PWF $(P,e)$ and $S:=N_P\cap X\subseteq X$. By {\bf Definition} \ref{df:hereditary-closure-finite-set} we have a sequence $(S_n, \sigma_n)_{n\in\NN}$. Since $S$ is finite, by {\bf Proposition} \ref{pr:hereditary-closure-finite-set}, we have a maximum set $S_m$ for the sequence $(S_n)_{n\in\NN}$. We denote by $\big(X_{(P,e)}, \sigma_{(P,e)}\big)$ the couple $\big(S_m, \sigma_m)$.
\end{definition}

\begin{definition}\label{df:binding-infinite-block-names}
  Let us consider a set of names $X\subseteq\NN$ and a PWF $(P,e)$ and consider $S:=\Fn(P)\cap X$. We define 
  \[
    (\nu X)(P,e):=(\nu X_{(P,e)})(P,e)\setminus X:= ((\nu X_{(P,e)})P_{\sigma_{(P,e)}}, e\setminus X)
  \]
\end{definition}

\begin{remark}\label{rk:general-nu-definition}
  The operator $(\nu X)$ for an infinite set is defined since $X_{(N_P,e)}$ is a finite set and $(Q,f){\setminus} X$ is defined for any~PWF $(Q,f)$ and any set $X$ ({\bf Definition} \ref{df:PWF-setminus-a-set}).
\end{remark}

\begin{example}
  Suppose that $x<y<z$ are names. Consider: $X:=\{x,t\}$, $P:=x^\uparrow(). y^\downarrow()$ and $e:= x\leftrightarrow y\leftrightarrow z; t\leftrightarrow u$. We compute ($(\nu X)(P, e)$ and for that we compute $X_{(P,e)}=\{x,y,z\}$. Then:
  \begin{align*}
    (\nu X)(P, e)&:= (\nu \{x,y,z\})(x^\uparrow(). y^\downarrow(), e)\setminus X&\equiv_\alpha (\nu z)(\nu y)(\nu x)(x^\uparrow(). y^\downarrow(), e)\setminus X\\
    &=(\nu z)(\nu y)(y^\uparrow(). y^\downarrow(), y\leftrightarrow z;t\leftrightarrow u)\setminus X&= (\nu z)(z^\uparrow(). z^\downarrow(), t\leftrightarrow u)\setminus X\\
    &=((\nu z)z^\uparrow().z^\downarrow, \Delta_\NN)&\\
  \end{align*}
\end{example}

\begin{lemma}\label{lm:computing-with-fusions}
  Let us consider fusions $e, f$ where $|e|\subseteq\NN^1$ and $|f|\subseteq\NN^2$ and define  $\sigma_\tau:=\displaystyle\prod_{x\in\NN^1}(x\leftrightarrow \tau(x))$ for  $\tau:\NN^1\to\NN^2$. Then:
  \begin{enumerate}
    \item\label{lm:computing-with-fusions-it:1} $ef\sigma_\tau\setminus\NN^1=\{(a,b)\in\NN^2\times\NN^2\ |\ \exists x_0,\dots x_{4k+1}\ \ a=x_0\ f x_1\ \tau\ x_2\ e\ x_3\ \tau\ x_4,\ \dots\ ,$
        \begin{flushright}
        $\dots, x_{4(k-1)}\ f\ x_{4(k-1)+1}\ \tau\ x_{4(k-1)+2}\ e\ x_{4(k-1)+3}\ \tau\  x_{4k}\ f\ x_{4k+1}\}$
      \end{flushright}
    \item\label{lm:computing-with-fusions-it:2} $ef\sigma_\tau\setminus\NN^1=e_\tau f$. 
  \end{enumerate}
\end{lemma}
\begin{proof}\ 
  \begin{enumerate}
    \item By definition of the transitive closure of $e\cup f\cup\{(x,\tau(x))\ x\in\NN^1\}$ we get the result. Observe that $a,b\in\NN^2$ because of the restriction $\setminus\NN^1$. 
    \item On the chain above observe that the blocks $x_{6j+1}\ \tau x_{6j+2}\ e\ x_{6j+3}$ can be replaced by $x_{6j+1}\ e_\tau\ x_{6j+3}$. This observation yields $ef\sigma_\tau\setminus\NN^1\subseteq e_\tau f$. The converse inclusion is straightforward since $e_\tau\subseteq ef\sigma_\tau\setminus\NN^1$.
  \end{enumerate} 
\end{proof} 

\begin{definition}\label{df:combinators-identity-and-parallel}
  We define the following fusions: $I:=\displaystyle\prod_{n\in\NN} (n.1\leftrightarrow n.2)$; $\Psi:=\displaystyle\prod_{n\in\NN}(n.1\leftrightarrow n.1.2)$ and $\Phi:=\Psi\ \ I^2=\displaystyle\prod_{n\in\NN}(n.1\leftrightarrow n.1.2)\displaystyle \prod_{n\in\NN}(n.1.2\leftrightarrow n.2.2)$. 
\end{definition}

\begin{corollary}\label{cr:composition-fusions-injections}
  Let us consider fusions $e, f$. Then:
  \begin{enumerate}
    \item\label{cr:composition-fusions-injections-it:1} $e^1f^2\Phi=\{(a,b)\ |\ \exists x_0, \dots, x_k \text{ s.t. } a=x_0\ \Phi x_1\ e^1\ x_2\ \Phi\ x_3\ f^2\ x_4\ \Phi \dots\ \Phi\ x_{k-2}\ e^1\ x_{k-1}\ \Phi\ x_k=b, k\in\NN^*\}$.
    \item\label{cr:composition-fusions-injections-it:2} $e^1 f^2\Phi\setminus \NN_1=e^{1.2} f^2\displaystyle\prod_{n\in\NN} (n.1.2\leftrightarrow n.2.2)=e^{1.2} f^2 I^2$.
    \item\label{cr:composition-fusions-injections-it:3} $e^1f^2 I\setminus \NN_1=e^2f^2$.
  \end{enumerate}
\end{corollary}
\begin{proof}\ 
  \begin{enumerate}
    \item By {\bf Lemma} \ref{lm:computing-with-fusions}.\ref{lm:computing-with-fusions-it:1}. 
    \item Applying {\bf Lemma} \ref{lm:computing-with-fusions}.\ref{lm:computing-with-fusions-it:2}: $e^1f^2\Phi\setminus\NN^1=e^1f^2I^2\Psi\setminus\NN^1=e^1f^2I^2\sigma_{\tau}\setminus\NN^1=(e^1)_\tau f^2I^2=e^{1.2}f^2I^2$ where \\ $\tau:\NN^1\to\NN^{1.2}, \tau(x.1):=x.1.2$. 
    \item Applying {\bf Lemma} \ref{lm:computing-with-fusions}.\ref{lm:computing-with-fusions-it:2}: $e^1f^2I\setminus\NN^1=e^1f^2\sigma_\mu\setminus\NN^1=(e^1)_\mu f^2=e^2f^2$ where $\mu:\NN^1\to\NN^2, \mu(x.1):=x.2$.
  \end{enumerate} 
\end{proof}

\begin{corollary}\label{cr:cr-composition-fusions-injections}
  Let us consider a fusion $e$. Then:
  \begin{enumerate}
    \item\label{cr:cr-composition-fusions-injections-it:1} $e^1\Phi=\{(a,b)\ |\ \exists x,y\ \ (a,x), (y,b)\in \Phi\text{ and } (x,y)\in e^1\}$. 
    \item \label{cr:cr-composition-fusions-injections-it:2}$e^1\Phi\setminus\NN_1= e^{1.2}\displaystyle\prod_{n\in\NN} (n.1.2\leftrightarrow n.2.2) = e^{1.2}I^2$.
    \item\label{cr:cr-composition-fusions-injections-it:3} $e^1I\setminus\NN_1=e^1\displaystyle\prod_{n\in\NN}(n.1\leftrightarrow n.2)\setminus\NN_1=e^2$. 
  \end{enumerate}   
\end{corollary}
\begin{proof}
  Apply the result above with $f:=\Delta_\NN$.
\end{proof}

\begin{proposition}\label{pr:nu-1-restriction-processes} 
  Let us consider $\tau:\NN^1\to\NN^2$ a substitution, (PWF) processes $(P,e), (Q,f)$ s.t. $\Fn(P,e)\subseteq \NN^1$ and $\Fn(Q,f)\subseteq \NN^2$. Let us define $\sigma_\tau:=\displaystyle\prod_{x\in\NN^1}(x\leftrightarrow \tau(x))$. Then $(\nu 1)(P|Q,ef\sigma_\tau)\equiv_\alpha (P_\tau|Q, e_\tau f)$.
\end{proposition}
\begin{proof}
  By {\bf Definition} \ref{df:binding-infinite-block-names} $(\nu 1)(P|Q, ef\sigma_\tau)\equiv_\alpha (P_\sigma|Q, ef\sigma_\tau\setminus \NN^1)$ for some $\sigma:S\subseteq \NN^1\to\NN^2$. By {\bf Lemma} \ref{lm:computing-with-fusions} $ef\sigma_\tau\setminus\NN^1=e_\tau f$. It remains to proof that $(P_\sigma |Q, e_\tau f)\equiv_\alpha (P_\tau |Q, e_\tau f)$. For the latter, by definition of $\equiv_\alpha$ on (PWF), it suffices to prove that $(P_\sigma|Q)_\rho = (P_\tau|Q)_\rho$ for $\rho(x):=x^\bullet_{e_\tau f}:=\min [x]_{e_\tau f}$. Recall that $\sigma(x)\simf{ef\sigma_\tau} x\simf{ef\sigma_\tau} \tau(x)$ for all $x\in S$ and, since $\sigma(x), \tau(x)\in\NN^2$, we have $\sigma(x)\simf{ef\sigma_\tau\setminus\NN^1} \tau(x)$, i.e.  $\sigma(x)\simf{e_\tau f} \tau(x)$ and then $\rho(\sigma(x))=\rho(\tau(x))$ for all $x\in S$, which ends the proof.
\end{proof}

\begin{definition}\label{df:constructors-on-PWF}\label{df:parallel-composition}
  Given $(P, e)$, $(Q, f)$ (PWF)-processes, $u, \vec{x}$ names and $\varepsilon$ a polarity, we define:
  \begin{itemize}
    \item $(P, e)\ \big|\ (Q, f):=(P|Q, ef)$. 
    \item $u^\varepsilon(\vec{x})(P, e):=(u^\varepsilon(\vec{x}).P, e)$ whenever $\forall x_i{\in}\vec{x}\quad [x_i]_e=\{x_i\}$ (otherwise this construction is not allowed). 
      %  \item $!(P, e):=(!P, e)$.  
    \item $\One:=(\one, \Delta_\NN)\in\overline{\Pi}$.
    \item Given a set $X$ and a (PWF) $(P, e)$ we have yet defined $(\nu X)(P, e)$. 
  \end{itemize}

  We get constructions as operators on (PWF), where those constructors are defined as follows: 

  \begin{tabular}{|l|l|}
    \hline
    {\sc Sets}&{\sc Constructors}\\
    \hline
    $\begin{array}{rcl}
      \mathcal{E}_{\vec{x}}&:=&\{e\in\mathcal{E}\ |\ \forall x_i{\in}\{\vec{x}\}\ \
      [x_i]_e{=}\{x_i\}\}\\[1ex]
        \NN^{<\omega}\times_\alpha \overline{\Pi}&:=&\{(\vec{x}, P, e){\in}
        \NN^{<\omega}\times\overline{\Pi}\ |\ e\in\mathcal{E}_{\vec{x}}\}\\[1ex]
          \overline{\Pi}_{\alpha}&:=&\{(\alpha.P,e)\ |\ \alpha\text{ is an action }, P{\in}\Pi\}\\
        \end{array}$&
        $\begin{array}{rcl}
          \One&\in&\overline{\Pi}\\
          \nu&:&\mathcal{P}(\NN)\times \overline\Pi\to\overline{\Pi}\\
          |&:&\overline{\Pi}\times\overline{\Pi}\to\overline{\Pi}\\
          u^\varepsilon&:&\NN^{<\omega}\times_\alpha\overline{\Pi}\to \overline{\Pi}\\
          %  !&:& \overline{\Pi}_\alpha\to \overline{\Pi}\\
        \end{array}$\\\hline
  \end{tabular}
\end{definition}

\subsection{(PWF) as a set generated by \texorpdfstring{$\Pi$}\ \ and \texorpdfstring{$\mathcal{E}$}\ \ under parallel composition}

Now we observe that there are immersions from $\Pi$ and $\mathcal{E}$ to $\overline{\Pi}$ which preserve the constructions of $\Pi$ and the join of $\mathcal{E}$:

\begin{definition}\label{df:immersions-over-PWF}
  Let us define $\gamma_1:\Pi\to\overline{\Pi}$ by means of $\gamma_1(P):=(P, \Delta_\NN)$ and $\gamma_2:\mathcal{E}\to\overline{\Pi}$ by means of $\gamma_2(e):=(\one, e)$. Let us define $\overline{\Pi}_0:=\operatorname{Im}(\gamma_1)$ and $\mathcal{E}_0:=\id(\gamma_2)$. 
\end{definition}

\begin{proposition}\label{pr:immersions-over-PWF}
  Given $P, Q\in\Pi$, $e,f\in\mathcal{E}$, $u, x$ names, $\vec{x}\in\mathcal{E}_{\vec{x}}$ a vector of names and $\varepsilon$ a polarity. 
  \begin{enumerate}
    \item $(P,e)=\gamma_1(P)|\gamma_2(e)$. Thus $\overline{\Pi}$ is generated from $\overline{\Pi}_0\cup\mathcal{E}_0$ by means of parallel composition.
    \item $\gamma_1(\one)=(\one, \Delta_\NN)=\One$.
    \item $\gamma_1(P)\ \big|\ \gamma_1(Q)=(P, \Delta_\NN)\ \big|\ (Q, \Delta_\NN)= (P|Q, \Delta_\NN)=\gamma_1(P|Q)$.
    \item $u^\varepsilon(\vec{x})(\gamma_1(P))=u^\varepsilon(\vec{x}).(P, \Delta_\NN)=(u^\varepsilon(\vec{x}).P, \Delta_\NN)=\gamma_1(u^\varepsilon(\vec{x}).P)$.
    \item $(\nu x)(\gamma_1(P))=(\nu x)(P, e)=((\nu x)P, \Delta_\NN)=\gamma_1((\nu x)P)$.
      %  \item $!(\gamma_1(\alpha.P))=!(\alpha.P, \Delta_\NN)=(!\alpha.P,\Delta_\NN)=\gamma_1(!\alpha.P)$.
    \item $\gamma_2(e)\ \big|\ \gamma_2(f)=(\one, e)\ \big|\ (\one, f)=(\one, ef)=\gamma_2(ef)$. 
  \end{enumerate}
\end{proposition}

Thus, identifying $P=\gamma_1(P)$ and $e=\gamma_2(e)$ we can write $(P,e)$ instead of $P|e$.

\begin{proposition}\label{pr:properties-constructions-on-PWF}
  We have the following properties:
  \begin{enumerate}
    \item $(\overline{\Pi}, |, \One, \equiv)$ is an abelian monoid.
    \item The structural equivalence on (PWF) respects the operators $\nu, |, !$ and $u^\varepsilon$, i.e.:
      \begin{center}
        \begin{prooftree}
          (P, e)\equiv (P',e)\quad  e\in\mathcal{E}_{\vec{x}}
          \justifies
          u^\varepsilon(\vec{x}).(P, e)\equiv u^\varepsilon(\vec{x}).(P',e)
        \end{prooftree}\hfill
        \begin{prooftree}
          (P,e)\equiv (P',e)
          \justifies
          (P,e)\ |\ (Q,f)\equiv (P',e)\ |\ (Q,f)
        \end{prooftree}\hfill
        \begin{prooftree}
          (P,e)\equiv (P',e)
          \justifies
          (\nu x)(P,e)\equiv (\nu x)(P',e)
        \end{prooftree}
      \end{center}
  \end{enumerate}
\end{proposition}
\begin{proof}
  The first statement is straightforward. For the second one:
  \begin{itemize}
    \item Suppose that $(P, e)\equiv (P',e)$, which implies that $P_{\sigma_e}\equiv P_{\sigma_e}'$ and then $(\sigma_e(u)^\varepsilon(\vec{x})).P_{\sigma_e}\equiv (\sigma_e(u)^\varepsilon(\vec{x})).P'_{\sigma_e}$ (observe that there is no capture of free names because $\sigma_e(z)=x_i$ iff $z=x_i$ for $x_i\in \vec{x}$). From that we get $(u^\varepsilon(\vec{x}).P, e)\equiv (u^\varepsilon(\vec{x}).P', e)$.
    \item From $(P, e)\equiv (P',e)$, we get $(P|Q, ef)\equiv (P'|Q', ef)$ because $P_{\sigma_{e}}\equiv P'_{\sigma_e}$ implies that $P_{\sigma_{ef}}\equiv P'_{\sigma_{ef}}$. 
      %we have $P_{\sigma_e}\equiv  P'_{\sigma_e}$ and then $!P_{\sigma_e}\equiv !P'_{\sigma_e}$ and we get $(!P,e)\equiv (!P',e)$. For the parallel composition is also straightforward.
    \item Suppose that $(P, e)\equiv (P',e)$, which implies that $P_{\sigma_e}\equiv P_{\sigma_e}'$ and let's try to prove that $(\nu x)(P,e)\equiv (\nu x)(P',e)$ which is equivalent to prove that $((\nu x)P\{x:=x^*_e\})_{\sigma_{e{\setminus}\{x\}}}\equiv((\nu x)P'\{x:=x^*_e\})_{\sigma_{e{\setminus}\{x\}}}$. Since $\sigma_{e{\setminus}\{x\}}(x)=x$, equivalently we have to prove that $(\nu x)P\sigma_{e{\setminus}\{x\}}\circ\{x:=x^*_e\}\equiv(\nu x)P'\sigma_{e{\setminus}\{x\}}\circ\{x:=x^*_e\}$. By {\bf Lemma} \ref{lm:sigma-e/sigma-e/x} it is equivalent to prove that $(\nu x)P\{x:=x^*_e\}\circ \sigma_e\equiv(\nu x)P'\{x:=x^*_e\}\circ\sigma_e$. We have two cases:
      \begin{itemize}
        \item If $x=x^*_e$, then we must prove that $(\nu x)P_{\sigma_e}\equiv (\nu x)P'_{\sigma_e}$, which is true since $P_{\sigma_e}\equiv P'_{\sigma_e}$.
        \item If $x\neq x^*_e$, then $x=x^\bullet_e$ and $x\notin \operatorname{Im}(\{x:=x^*_e\}\circ \sigma_e)$ and we conclude that $x\notin \Fn(P\{x:=x^*_e\}\circ \sigma_e)$ and $x\notin \Fn(P'\{x:=x^*_e\}\circ \sigma_e)$. We get $(\nu x)P\{x:=x^*_e\}\circ \sigma_e\equiv P\{x:=x^*_e\}\circ \sigma_e=P_{\sigma_e}\{x:=x^*_e\}\overset{(a)}{\equiv} P'_{\sigma_e}\{x:=x^*_e\}\equiv (\nu x)P'\{x:=x^*_e\}\circ\sigma_e$ where the equivalence $(a)$ is a consequence of $P_{\sigma_e}\equiv P'_{\sigma_e}$. 
      \end{itemize}
  \end{itemize}
\end{proof}

\subsection{Parallel composition in terms of the left adjoint of implication}
Remember {\bf Definition}\ref{df:iota-immersions}, where we fix injections  $\iota_1, \iota_2:\NN\to\NN$ such that $\iota_1(\NN)\cap\iota_2(\NN)=\emptyset$ and $\iota_1(\NN)\cup \iota_2(\NN)=\NN$. More precisely we take  $\iota_1(n):=2n+1$ and $\iota_2(n):=2n$. For $i=1,2$ we denote $(P,e)^i$ the process $(P,e)_{\iota_1}=(P_{\iota_i}, e_{\iota_i})$, i.e. the result of applying $\iota_i$ to the whole names of the PWF $(P,e)$. We also denote as $(P,e)^{-i}$ the process $(P, e)_{\iota_i^{-1}}$, which has meaning only if $\Fn(P,e)\subseteq \NN^i$.  

\begin{definition}\label{df:operations-with-processes}
  
  Define the following operations:
  \[
    \begin{array}{rcl}
      (P,e)\bullet (Q,f)&:=&(P,e)^1|(Q,f)^2, for i=1,2\\
      (P,e)*_i (Q,f)&:=&((\nu_i)((P,e)|(Q,f)^i))^{(3-i)^{-1}}\\
    \end{array}
  \]
\end{definition}

Observe that for $i=1$, $3-i=2$ and for $i=2$, $3-i=1$.  $\Fn((\nu_i)((P,e)|(Q,f)^i)\subseteq \NN^{3-i}$ and thus we can apply $\iota_{3-i}^{-1}$ to this process. 

On the polarized $\pi$-calculus \cite{Beffara-thesis} we have an expression of the parallel constructor $|$ in terms of the left adjoint $*_1$ of $\multimap$. Let us consider polarized processes $P, Q$ and define the combinator \[\Phi(P,Q):=\displaystyle\prod_{n\in\Fn(P,Q)}\big(n.1\leftrightarrow n.1.2\leftrightarrow n.2.2\big)\] 
It is no difficult to prove --and we will do that-- $\Phi(P, Q)*_1 P=\Big((\overline\nu_1)\big(P^1\ |\ \displaystyle\prod_{n\in\Fn(P, Q)}\big(n.1\leftrightarrow n.1.2\leftrightarrow n.2.2\big)\Big)^{-2}\equiv_\alpha \Big(P^{1.2}|\displaystyle\prod_{n\in\Fn(P, Q)}\big(n.1.2\leftrightarrow n.2.2\big)\Big)^{-2}=\Big(P^1|\displaystyle\prod_{n\in\Fn(P, Q)}\big(n.1\leftrightarrow n.2\big)\Big)$. Then, since $\equiv_\alpha$ is a congruence, $\Phi(P, Q)*_1 P*_1 Q\equiv_\alpha \Big(P^1|\displaystyle\prod_{n\in\Fn(P, Q)}\big(n.1\leftrightarrow n.2\big)\Big)*_1 Q =(\overline{\nu}_1)\Big(P^1|Q^1|\displaystyle\prod_{n\in\Fn(P, Q)}\big(n.1\leftrightarrow n.2\big)\Big)^{-2}\equiv_\alpha (P^2|Q^2)^{-2}=P|Q$. 

However, since on (PWF) terms we have infinitely many fusions, we can define a single combinator (in facto, a fusion) $\Phi$ to express parallel composition in terms of $*_1$:

\begin{lemma}\label{lm:bisimilarity-and-substitution}
  Let us consider a (PWF) $(P, e)$. Then:
  \begin{enumerate}
    \item Let us consider a substitution $\sigma:X\to Y$ where $X\cap Y=\emptyset$ and a fusion $f:=\displaystyle\prod_{x\in X}(x\leftrightarrow \sigma(x))$. 

      Then $(\nu X)(P,e f)=((\nu X_{(P,e)})P_{\sigma_{(P,e f)}}, e)$.
    \item $(\nu_1)(P^1|Q^2, e^1f^2\Phi)\equiv_\alpha (P^{1.2}|Q^2, e^1f^2\Phi\setminus\NN^1)\equiv_\alpha (P^{1.2}|Q^2, e^{1.2}f^2I^2)$. 
    \item $(\nu_1)(P^1|Q^2, e^1f^2 I)\equiv_\alpha,(P^2|Q^2, e^1f^2 I\setminus\NN^1)\equiv_\alpha (P^2|Q^2, e^2f^2)$.
  \end{enumerate}
\end{lemma}   
\begin{proof}\ 
  \begin{enumerate}
    \item By {\bf Definition} \ref{df:binding-infinite-block-names} $(\nu X)(P,e f)=((\nu X_{(P,e f)})P_{\sigma_{(P,e f)}}, e f \setminus X)=((\nu X_{(P,e f})P_{\sigma_{(P,e f)}}, e)$. 
    \item We have $(\nu_1)(P^1|Q^2, e^1f^2\Phi) =\big(\nu \NN^1_{(P, e^1f^2\Phi)}\big)(P^1|Q^2, e^1f^2\Phi)\setminus\NN^1 = ((\nu \NN^1_{P,e^1f^2\Phi})(P^1|Q^2)_{\sigma_{(P,e^1f^2\Phi)}}, e^1f^2\Phi\setminus\NN^1)$. 

      By {\bf Proposition} \ref{pr:hereditary-closure-finite-set} $\sigma_{(P, e^1f^2\Phi)}:\NN^1_{(P, e^1f^2\Phi)}\to \NN^2$ and $\sigma_{(P, e^1f^2\Phi)}(n.1)=n.2$. Thus we get $(\nu_1)(P^1|Q^2, e^1f^2\Phi)=(P^2|Q^2, e^1f^2\Phi\setminus\NN^1)\equiv_\alpha (P^2|Q^2, e^{1.2}f^2 I^2)$, the last by {\bf Corollary} \ref{cr:composition-fusions-injections}.\ref{cr:composition-fusions-injections-it:2}.
    \item Analogously, $(\nu_1)(P^1|Q^2, e^1f^2I)=((P^1|Q^2)_{\sigma_{(P,e^1f^2I)}},e^1f^2I)\setminus\NN^1=(P^2|Q^2, e^2f^2I\setminus\NN^1)=(P^2|Q^2, e^2f^2)$, again by {\bf Proposition} \ref{pr:hereditary-closure-finite-set} and the last step by {\bf Corollary} \ref{cr:composition-fusions-injections}.\ref{cr:composition-fusions-injections-it:3}.
  \end{enumerate}
\end{proof}

\begin{proposition}\label{pr:parallel-comp-in-terms-adj}
  Given a (PWF) $(P, e)$, $(Q, f)$, we have that $\Phi *_1 (P, e) *_1 (Q, f)\equiv_\alpha (P,e)\ |\ (Q,f)$
\end{proposition} 
\begin{proof}
  \begin{align*}
    \Phi *_1 (P, e)
    &:= \big((\nu_1)(P^1|e^1\Phi)\big)^{-2} &\text{ by definition of }*_1\\
    &\equiv_\alpha(P^{1.2}, e^{1.2}I^2)^{-2} & \text{by {\bf Lemma} \ref{lm:bisimilarity-and-substitution}}\\
    &=(P^1, e^1I)& \text{applying } \iota_2^{-1}\\
  \end{align*}
  Then we get
  \begin{align*}
    \Phi *_1 (P, e) *_1 (Q, f)&\equiv_\alpha (P^1, e^1I)*_1 (Q, f) & \text{ by the remark above }\\
    & := (\nu _1)(P^1|Q^1, e^1f^1I)^{-2}& \text{ by definition of }*_1\\
    &\equiv_\alpha (P^2|Q^2, e^2f^2)^{-2} & \text{ by {\bf Lemma} } \ref{lm:bisimilarity-and-substitution}\\
    &=(P|Q, ef) = (P,e)|(Q,f)& \text{ applying } \iota_2^{-1}
  \end{align*}
\end{proof}   

\subsection{Poles, polarities, orthogonality and behaviours} 
The notion of \emph{poles} is central in \emph{Classical Realizability} \cite{Krivine94}, \cite{Krivine03} and is maintained on the algebraic presentations \cite{Ferrer-et-al}, \cite{Miquey-thesis} and \cite{Miquel-IA}. These poles are parameters for the Realizability definition and allow to give a non-trivial interpretation for negation, unlike what is done on Intuitionistic Realizability \cite{Kleene45}, \cite{Kreisel52}. The interpretation of formul\ae\ is strongly inspired by the phase semantics \cite{Girard87}. 

We start by considering poles that are barely closed under $\alpha$-equivalence and doing the part of the theory that does not require any form of (anti)reduction closure. As is done in the semantics of phases, we interpret the language in the set of \emph{behaviors}, which are the sets $X$ closed under double orthogonal (i.e.: $X=X^{\perp\perp}$). The fundamental connective that we will use is the multiplicative conjunction $\otimes$. Intuitively, $X\otimes Y$ corresponds to (by means of renaming) forcing processes of $X$ not to communicate with those of $Y$. As we need to get a behaviour, we close it by double orthogonal.    

Using the orthogonality defined by the poles and the defined notion $\otimes$, we then introduce the connectives that correspond to the Multiplicative Linear Logic (MLL).
\begin{definition}\label{df:poles-orthogonality}
  A \emph{pole} is a set of closed processes $\bbot\subseteq \overline{\Pi}_0:=\{(P,e)\in\overline{\Pi}\ |\ \Fn(P,e)=\emptyset\}$ that is closed under $\equiv_\alpha$, i.e.: if $(P,e)\equiv_\alpha (Q,e)\in\bbot$ then $(P,e)\in\bbot$. 
  
  Given (PWF)-processes $(P,e), (Q,f)$ we say that they are \emph{orthogonal} iff $(\overline{\nu})\big((P,e)|(Q,f)\big):=(\overline{\nu})(P|Q, ef)\in\bbot$. We write $(P,e)\perp (Q,f)$ to abbreviate that $(P,e)$ and $(Q,f)$ are orthogonal.

  Given $\bbot$ we define an orthogonal operator $(\ )^\perp{:}\mathcal{P}(\overline{\Pi}){\to}\mathcal{P}(\overline{\Pi})$ by  $A^\perp:=\{(P,e)\in\overline{\Pi}\ |\ \forall(Q,f){\in}A\ \ (P,e){\perp}(Q,f)\}$.

  We define $\mathbb{P}:=\mathcal{P}(\Pi)$ and the set of \emph{behaviours} by means of $\mathbb{B}:=\{A\in \mathcal{P}(\overline{\Pi})\ |\ A^{\perp\perp}=A\}\subseteq \mathbb{P}$. 
\end{definition}

\begin{proposition}\label{pr:lattices}
  Let us consider the structures $\mathcal{P}:=(\mathbb{P}, \bigcap, \bigcup, \subseteq)$ and $\mathcal{B}:=(\mathbb{B}, \bigcap, \bigvee, \subseteq)$ where $\bigvee\mathfrak{B}:= \Big(\bigcup \mathfrak{B}\Big)^{\perp\perp}$ for all $\mathfrak{B}\subseteq \mathbb{B}$. Then:
  \begin{enumerate}
    \item $\mathcal{P}:=(\mathbb{P}, \bigcap, \bigcup, \subseteq)$ is a complete lattice with top element $\overline\Pi$ and bottom element $\emptyset$.
    \item $\mathcal{B}:=(\mathbb{B}, \bigcap, \bigvee, \subseteq)$ is a complete lattice with top element $\overline\Pi$ and bottom element $\emptyset^{\perp\perp}$. 
  \end{enumerate}
\end{proposition}

\begin{proposition}\label{pr:poles-orthogonality}
  For all $A, B\subseteq\mathcal{P}(\Pi)$ and for all $\mathfrak{B}\subseteq \mathcal{P}(\Pi)$ we have:
  \begin{enumerate}
    \item \label{it01:poles-orthogonality}If $A\subseteq B$ then $B^\perp\subseteq A^\perp$.
    \item\label{it02:poles-orthogonality} $A\subseteq A^{\perp\perp}$ and $A^\perp=A^{\perp\perp\perp}$.
    \item\label{it03:poles-orthogonality} $\Big(\bigcup_{X\in\mathfrak{B}}X\Big)^\perp=\bigcap_{X\in\mathfrak{B}} X^\perp$ and if $\mathfrak{B}\neq\emptyset$ then $\Big(\bigcap_{X\in\mathfrak{B}}X\Big)^\perp \supseteq \Big(\bigcup_{X\in\mathfrak{B}}X^\perp\Big)$ while the reverse inclusion is in general false. 
    \item\label{it04:poles-orthogonality} $\Big(\bigcup_{X\in\mathfrak{B}}X^{\perp\perp}\Big)^\perp= \Big(\bigcup_{X\in\mathfrak{B}}X\Big)^\perp$. 
  \end{enumerate}
\end{proposition}
\begin{proof}\ 
  \begin{enumerate}
    \item[1.,2.,3.] Evident by definition. Observe that $(\overline\Pi, \overline\Pi, \perp =\{((P,e), (Q,f))\ |\ (P,e)\perp (Q,f)\})$ is called a \emph{polarity} on \cite{Ferrer-et-al} and it defines a Galois connextion \cite{Birkhoff}.
    \item[4.] $\Big(\bigcup_{X\in\mathfrak{B}}X^{\perp\perp}\Big)^\perp=\bigcap_{X\in\mathfrak{B}} X^{\perp\perp\perp}=\bigcap_{X\in\mathfrak{B}} X^\perp=\Big(\bigcup_{X\in\mathfrak{B}}X\Big)^\perp$.  
  \end{enumerate}
\end{proof}

\begin{definition}\label{df:operations-with-truth-values}
  We define the following operations on $\mathbb{P}:=\mathcal{P}(\overline{\Pi})$:
  \[ 
    \begin{array}{rcl}
      A\ |\ B&:=&\{(P,e)|(Q,f)\ |\ (P,e)\in A,\ (Q,f)\in B\}\\
      A\bullet B&:=&\{(P,e)\bullet(Q,f)\ |\ (P,e)\in A,\ (Q,f)\in B\}\\
      A\ \|\ B&:=&(A\ |\ B)^{\perp\perp}\\
      A*_i B&:=&\{(P,e)*_i(Q,f)\ |\ (P,e)\in A,\ (Q,f)\in B\}^{\perp\perp} (for i=1 or i=2)\\
      A\otimes B&:=&(A\bullet B)^{\perp\perp}\\
      A\parr B&:=&(A^\perp\otimes B^\perp)^\perp\\
      A\multimap B&:=&(A\otimes B^\perp)^\perp\\
      {\bf 1}&:=&\{(\one, \Delta_\NN)\}^{\perp\perp}. 
    \end{array}
  \]
\end{definition}

\begin{proposition}\label{pr:operations-with-truth-values}
  Let us consider $A, B, C\in\mathbb{P}$. Then:
  \begin{enumerate}
    \item\label{it01:operations-with-truth-values} $C\subseteq (A\bullet B)^\perp\Longleftrightarrow C*_1 A\subseteq B^\perp \Longleftrightarrow C*_2B\subseteq A^\perp$.
    \item\label{it02:operations-with-truth-values} $(A\bullet B)^\perp = (A^{\perp\perp}\bullet B^{\perp\perp})^\perp$. 
    \item\label{it03:operations-with-truth-values} $C\subseteq A\multimap B\Longleftrightarrow C*_1 A\subseteq B^{\perp\perp}\Longleftrightarrow C*_2B^\perp\subseteq A^\perp$.
    \item\label{it04:operations-with-truth-values} If $B\in\mathbb{B}$ then $C\subseteq A\multimap B\Longleftrightarrow C*_1 A\subseteq B$.
    \item\label{it05:operations-with-truth-values} $A\otimes B=A^{\perp\perp} \otimes B^{\perp\perp}$, $A\parr B=A^{\perp\perp} \parr B^{\perp\perp}$ and $A\multimap B=A^{\perp\perp}\multimap B^{\perp\perp}$.
    \item\label{it06:operations-with-truth-values} $C*_1A=\bigcap \{B\in\mathbb{P}\ |\ C\subseteq A\multimap B\}=\min\{B\in\mathbb{P}\ |\ C\subseteq A\multimap B\}$. 
  \end{enumerate}
\end{proposition}
\begin{proof}\ 
  \begin{enumerate}
    \item
      \[
        \begin{array}{rcccclcl}
          C\subseteq (A\bullet B)^\perp&\Leftrightarrow&\forall p\in C&\forall q\in A& \forall r\in B & p\perp q\bullet r&&\\
                                       &\Leftrightarrow&\forall p\in C&\forall q\in A& \forall r\in B &(\overline{\nu})(p|q^1|r^2)\in\bbot&&\\
                                       &\Leftrightarrow&\forall p\in C&\forall q\in A& \forall r\in B & (\overline{\nu}_2)\Big((\overline{\nu}_1)(p|q^1)|r^2\Big)\in\bbot&&\\
                                       &\Leftrightarrow&\forall p\in C&\forall q\in A& \forall r\in B &
          (\overline{\nu})\Big((\overline{\nu}_1)(p|q^1)^{-2}|r\Big)\in\bbot&&\\
                                                                            &\Leftrightarrow&\forall p\in C&\forall q\in A& \forall r\in B &
          p*_1q\perp r&\Leftrightarrow&C*_1A\subseteq B^\perp\\
        \end{array}
      \]
      The proof of $C\subseteq (A\bullet B)^\perp\Longleftrightarrow C*_2B\subseteq A^\perp$ is --\emph{mutatis mutantis}-- the same.
    \item
      \[ 
        \begin{array}{rclccccl}
          (A\bullet B)^\perp\subseteq (A\bullet B)^\perp&\Leftrightarrow&(A\bullet B)^\perp *_1A&\subseteq& B^\perp=B^{\perp\perp\perp}&&&\\
                                                        &\Leftrightarrow&(A\bullet B)^\perp&\subseteq&(A\bullet B^{\perp\perp})^\perp&&\\
                                                        &\Leftrightarrow&(A\bullet B)^\perp *_2 B^{\perp\perp}&\subseteq& A^\perp=A^{\perp\perp\perp}& \Leftrightarrow&(A\bullet B)^\perp\subseteq (A^{\perp\perp}\bullet B^{\perp\perp})^\perp\\
        \end{array}
      \]
      By monotony of $\bullet$ and the closure $()^{\perp\perp}$ we get $A\bullet B\subseteq A^{\perp\perp}\bullet B^{\perp\perp}$. Taking one more $()^\perp$ we obtain $(A\bullet B)^\perp\supseteq (A^{\perp\perp}\bullet B^{\perp\perp})^\perp$, which ends the proof.
    \item Apply \ref{pr:operations-with-truth-values}.\ref{it01:operations-with-truth-values}, using that $A\multimap B=(A\bullet B^\perp)^\perp$.  
    \item A direct consequence of \ref{pr:operations-with-truth-values}.\ref{it03:operations-with-truth-values}. 
    \item Apply \ref{pr:operations-with-truth-values}.\ref{it02:operations-with-truth-values} and the definitions of $\otimes$, $\parr$ and $\multimap$ using $\bullet$.
    \item Applying \ref{pr:operations-with-truth-values}.\ref{it03:operations-with-truth-values} we have that $C*_1A\subseteq C*_1 A\Longleftrightarrow C\subseteq A\multimap (C*_1A)$. Then $C*_1A\in\{B\ |\ C\subseteq A\multimap B\}$. On the other hand, if $C\subseteq A\multimap B$, then $C*_1A\subseteq B$. 
  \end{enumerate}
\end{proof}

\begin{lemma}\label{lm:otimes-vee-compatibility}
  Let us consider  $\mathfrak{B}\subseteq \mathcal{P}(\mathbb{B})$ and $A\in\mathbb{B}$. Then $A\otimes \bigvee\mathfrak{B}=\bigvee_{B\in\mathfrak{B}}(A\otimes B)$.
\end{lemma}
\begin{proof}
  \[
    \begin{array}{rclclc}
      A\otimes\bigvee_{B\in\mathfrak{B}} B & \underset{\text{def of }  \vee}{=} & A\otimes \Big(\bigcup_{B\in\mathfrak{B}}B\Big)^{\perp\perp} & \underset{\text{\ref{pr:operations-with-truth-values}.\ref{it05:operations-with-truth-values}}}{=}&A\otimes  \bigcup_{B\in\mathfrak{B}}B&\underset{\text{def of }\otimes}{=}\\
      \Big(A\bullet\bigcup_{B\in \mathfrak{B}}B\Big)^{\perp\perp} & = & \Big(\bigcup_{B\in\mathfrak{B}}(A\bullet B)\Big)^{\perp\perp} &
      \underset{\text{\ref{pr:poles-orthogonality}.\ref{it04:poles-orthogonality}}}{=}& \Big(\bigcup_{B\in\mathfrak{B}}(A\bullet B)^{\perp\perp}\Big)^{\perp\perp} &
      \underset{\text{def of }\otimes}{=}\\
      \Big(\bigcup_{B\in\mathfrak{B}}(A\otimes B)\Big)^{\perp\perp} &  \underset{\text{def of }\vee}{=}&{\bigvee_{B\in\mathfrak{B}}}(A\otimes B)&&&\\
    \end{array}
  \]
\end{proof}

\begin{lemma}\label{lm:compatibility-parallel-join}
  Let us consider $\mathfrak{B}\subseteq \mathbb{P}=\mathcal{P}(\overline\Pi)$ and $A\subseteq \mathbb{P}$. Then $\left(\bigvee \mathfrak{B}\right) | A \subseteq \bigvee \{B|A\ |\ B\in\mathfrak{B}\}=\left((\bigcup \mathfrak{B}) | A\right)^{\perp\perp}$. 
\end{lemma}
\begin{proof}
  Let us consider $(P,e)\in \bigvee \mathfrak{B}$, $(Q, f)\in A$ and $(H,h)\in \left((\bigcup \mathfrak{B})|A\right)^\perp$. It suffices to prove that $\overline\nu (P|Q|H, efh)\in\bbot$. By the hypothesis that  $(P,e)\in \bigvee \mathfrak{B}$ it suffices to prove that $(Q|H, fh)\in \left(\bigcup \mathfrak{B}\right)^\perp$. Let us consider that $(R,r)\in \bigcup\mathfrak{B}$. Then, it suffices to prove that $\overline\nu (H|R|Q, hrq)\in\bbot$. This sentence is true because $(R|Q, rf)\in \left(\bigcup \mathfrak{B}\right)|A$ and $(H,h)\in \left((\bigcup \mathfrak{B})|A\right)^\perp$.
\end{proof}

\subsection{Adequacy Lemma} 
We define the Realizability relation on behaviours through: 
\begin{definition}\label{df:Realizability-on-behaviours} 
Given a pole $\bbot$, a behaviour $A\in\mathbb{B}$ and a (PWF) $P, e)$, we say that $(P, e)$ \emph{realizes} $A$ iff $(P, e)\in A$. We will indistincly use the notations $(P, e)\Vdash A$ or $(P, e)\in A$. 
\end{definition}
As usually on Realizability, we will interpret the language of logic (in this case of MLL) as sets taken from some set whose elements have computational meaning. In our case we interpret the language on behaviours. Given a pole $\bbot$, the theory of the corresponding \emph{Realizability model} is the set of formul\ae\ whose interpretation is realized. 

The result which establishes that Realizability is correctly defined is the \emph{Adequacy Lemma}, which consists in proving that the rules of logic (in this case the rules of MLL) deduce true consequences from true premises. In the remainder of the section we prove all the technical lemmas to prove the Adequacy Lemma, which will be done on the next section.
\begin{lemma}\label{lm:identityt-adequacy}
  Let us consider $A\subseteq \mathcal{P}(\overline\Pi)$. Then $I\in A^\perp\parr A$.  
\end{lemma}
\begin{proof}
  $A^\perp\parr A=(A^{\perp\perp}\otimes A^\perp)^\perp=(A\bullet A^\perp)^\perp$, the last equality by {\bf Proposition} \ref{pr:operations-with-truth-values} and {\bf Definition} \ref{df:operations-with-truth-values}. Let us consider $(P,e)\in A$ and $(Q,f)\in A^\perp$, which implies that $(\overline\nu)(P|Q, ef)\in\bbot$. Let us consider $(P,e)\bullet (Q,f)=(P^1|Q^2, e^1 f^2)\in (A\bullet A^\perp)$. 

  It is enough to prove that $(\one, I)\bbot (P^1|Q^2, e^1 f^2)$, which is equivalent to  $(\overline{\nu})(\one|P^1|Q^2, e^1f^2 I)\in\bbot$. By {\bf Lemma} \ref{lm:bisimilarity-and-substitution} we have that $(\overline\nu)(\one|P^1|Q^2, e^1 f^2 I)\equiv_\alpha (\nu_2)(\nu_1)(P^1|Q^2, e^1 f^2 I)\equiv_\alpha (\nu_2)(P^2|Q^2, e^2 f^2)\equiv_\alpha (\overline\nu)(P|Q, ef)$ being the last a member of $\bbot$. 
\end{proof}   

\begin{lemma}\label{lm:asociativity-adequacy}
  Let us consider $A, B, C\subseteq\mathcal{P}(\overline{\Pi})$. Then $\sigma_\tau\in A\otimes (B\otimes C)\multimap (A\otimes B)\otimes C$ for some $\tau:\NN^1\to\NN^2$.
\end{lemma}   
\begin{proof}
  Let us define $\tau:\NN^1\to\NN^2$ by the following sentences:
  \begin{align*}
    \tau(n.1.1)&:=n.1.1.2\\
    \tau(n.1.2.1)&:=n.2.1.2\\
    \tau(n.2.2.1)&:=n.2.2
  \end{align*}
  Observe that the definition is correct since $\NN^1=\NN^{1.1}\biguplus \NN^{1.2.1}\biguplus\NN^{2.2.1}$ and that $\tau(x)\in\NN^2$ for all $x\in\NN^1$.

  We have the following identities: 
  \[
    A\otimes (B\otimes C)\multimap (A\otimes B)\otimes C=\Big((A\otimes (B\otimes C))\otimes ((A\otimes B)\otimes C)^\perp\Big)^\perp=\Big((A\bullet (B\bullet C))\bullet((A\bullet B)\bullet C)^\perp\Big)^\perp 
  \]
  Then a generic process of $(A\bullet (B\bullet C))\bullet((A\bullet B)\bullet C)^\perp$  has the shape $(P^{1.1}, e^{1.1})|(Q^{1.2.1}, f^{1.2.1})|(R^{2.2.1}, g^{2.2.1})|(H^2, h^2)=(P^{1.1}|Q^{1.2.1}|R^{2.2.1}|H^2, e^{1.1}f^{1.2.1}g^{2.2.1}h^2)$ where: 
  \begin{align*}
    (P,e)&\in A\\
    (Q,f)&\in B\\
    (R,g)&\in C\\
    (H,h)&\in ((A\bullet B)\bullet C)^\perp
  \end{align*}
  It suffices to prove that \begin{equation}\label{eq:asociativity-adequacy}(\nu_2)(\nu_1)(P^{1.1}|Q^{1.2.1}|R^{2.2.1}|H^2, e^{1.1}f^{1.2.1}g^{2.2.1}h^2\sigma_\tau)\in\bbot\end{equation}. 

  On the other hand, we have that $(P^{1.1}|Q^{2.1}|R^2, e^{1.1}f^{2.1}g^2)\in ((A\bullet B)\bullet C)$ and then \[\overline{\nu}(P^{1.1}|Q^{2.1}|R^2|H, e^{1.1}f^{2.1}g^2h)\equiv (\nu_2)(P^{1.1.2}|Q^{2.1.2}|R^{2.2}|H^2, e^{1.1.2}f^{2.1.2}g^{2.2}h^2)\in\bbot\] 

  Applying {\bf Proposition} \ref{pr:nu-1-restriction-processes} we have $(\nu_1)(P^{1.1}|Q^{1.2.1}|R^{2.1.2}|H^2, e^{1.1}f^{1.2.1}g^{2.1.2}h^2\sigma_\tau)\equiv_\alpha$\\ 
  $\Big((P^{1.1})_\tau|(Q^{1.2.1})_\tau|(R^{2.2.1})_\tau|H^2, (e^{1.1})_\tau (f^{1.2.1})_\tau (g^{2.2.1})_\tau h^2\Big)=\big(P^{1.1.2}|Q^{2.1.2}|R^{2.2}|H^2, e^{1.1.2}f^{2.1.2}g^{2.2}h^2\big)$ which proves the sentence (\ref{eq:asociativity-adequacy}), thus ending the proof.  
\end{proof}

\begin{lemma}\label{lm:asociativity-adequacy-3}  
  Let us consider $A, B, C\subseteq\mathcal{P}(\overline{\Pi})$. Then $\sigma_{\tau}\in (A\otimes B)\otimes C\multimap A\otimes (B\otimes C)$ for some $\tau:\NN^1\to\NN^2$.
\end{lemma}
\begin{proof}
  Same proof as for the {\bf Lemma} \ref{lm:asociativity-adequacy}, but with $\tau$ given by $\tau(n.1.1):=n.1.2$, $\tau(n.2.1.1):=n.1.2.2$ and $\tau(n.2.1):=n.2.2.2$.  
\end{proof}

\begin{lemma}\label{lm:asociativity-adequacy-2}  
  Let us consider $A, B\subseteq\mathcal{P}(\overline{\Pi})$. Then $\sigma_{\tau}\in (A\otimes B)\multimap (B\otimes A)$ for some $\tau:\NN^1\to\NN^2$.
\end{lemma}
\begin{proof}
  Same proof as for the {\bf Lemma} \ref{lm:asociativity-adequacy}, but with $\tau$ given by $\tau(n.1.1):=n.2.2$, $\tau(n.2.1):=n.1.2$.  
\end{proof}

\begin{lemma}\label{lm:asociativity-adequacy-4}
  Let us consider $A\in\mathcal{P}(\overline{\Pi})$. Then $A\multimap ({\bf 1}\otimes A)$, $({\bf 1}\otimes A)\multimap A$, $A\multimap (A\otimes {\bf 1})$ and $(A\otimes {\bf 1})\multimap A$ are inhabited by fusions of the shape $\sigma_\tau$ for suitable substitutions $\tau$. 
\end{lemma}
\begin{proof}
  Similar to the proof of {\bf Lemma} \ref{lm:asociativity-adequacy}. For instance, consider $\tau:\NN^1\to\NN^{2.2}$ by $\tau(n.1):=n.2.2$ and prove that $\sigma_\tau=\displaystyle\prod_{n\in\NN}(n.1\leftrightarrow n.2.2)\in A\multimap ({\bf 1}\otimes A)$. Let us consider $(H,h)\in ({\bf 1}\otimes A)^\perp$ and $(P,e)\in A$. We must prove that $\overline\nu(P^1|H^2, e^1h^2\sigma_\tau)\in\bbot$. Again by {\bf Proposition} \ref{pr:nu-1-restriction-processes} this process is $\alpha$-equivalent to $(\nu_2)(P^{2.2}|H^2, e^{2.2}h^2)\equiv_\alpha\overline\nu(H|P^2, h\ e^2)$. The last is in $\bbot$ since $(P^2, e^2)\in{\bf 1}\otimes A$ and $(H,h)\in({\bf 1}\otimes A)^\perp$. 
\end{proof}

\begin{lemma}\label{lm:composition-linear-arrow}
  There exists a fusion of the shape $\sigma_\tau$ for some substitution $\tau$ s.t. $\sigma_\tau\in (A\multimap B)\multimap (B\multimap C)\multimap A\multimap C$
\end{lemma}   
\begin{proof}
  Let us consider the substitution $\tau:\NN^{1.2.2}\biguplus \NN^{2.1}\biguplus \NN^{2.1.2}\to \NN^{1.1}\biguplus\NN^{1.1.2}\biguplus\NN^{2.2.2}$ given by the sentences $\tau(n.1.2.2):=n.1.1$, $\tau(n.2.1):=n.1.1.2$ and $\tau(n.2.1.2):=n.2.2.2$. Observe that $\NN^{1.2.2}$ $\NN^{2.1}$ $\NN^{2.1.2}$ $\NN^{1.1}$ $\NN^{1.1.2}$ $\NN^{2.2.2}$ is a partition of $\NN$.

  By definition of $\multimap$ we have $(A\multimap B)\multimap (B\multimap C)\multimap A\multimap C=\Big((A\bullet B^\perp)^\perp\bullet \big((B\bullet C^\perp)^\perp\bullet (A\bullet C^\perp)\big)\Big)^\perp$. Let us consider $(P, e)\in(A\bullet B^\perp)^\perp$, $(Q, f)\in (B\bullet C^\perp)^\perp$, $(R, g)\in A$ and $(S,h)\in C^\perp$. To prove the result, it is enough to show that $\overline\nu \Big(Q^{1.2}|P^1|R^{1.2.2}|S^{2.2.2}, \sigma_\tau f^{1.2}e^1 g^{1.2.2} h^{2.2.2}\Big)\in\bbot$. Let us split the bind $\overline\nu$ as \[\overline\nu=(\nu\NN^{2.2.2})(\nu\NN^{2.1.2})(\nu\NN^{1.1.2})(\nu\NN^{2.1})(\nu \NN^{1.1})(\nu \NN^{1.2.2})\] 

  By hypothesis $(\nu_1)(P|R^1, eg^1)\in (B^{\perp\perp})^2=B^2$. Applying $\iota_1$ we get $(\nu_{1.1})(P^1|R^{1.1}, e^1g^{1.1})\in B^{2.1}$. Generalizing the {\bf Proposition} \ref{pr:nu-1-restriction-processes} we have $(\nu_{1.2.2})\Big(P^1|R^{1.2.2}, e^1 g^{1.2.2} \displaystyle\prod_{n\in\NN}(n.1.2.2\leftrightarrow n.1.1)\Big)\equiv_\alpha \Big(P^1|R^{1.1}, e^1g^{1.1}\Big)$ and then $(\nu_{1.1})(\nu_{1.2.2})\Big(P^1|R^{1.2.2}, e^1 g^{1.2.2} \displaystyle\prod_{n\in\NN}(n.1.2.2\leftrightarrow n.1.1)\Big)\equiv_\alpha (\nu_{1.1})\Big(P^1|R^{1.1}, e^1g^{1.1}\Big)\in B^{2.1}$. 

  In particular, $(\nu_{1.1})(\nu_{1.2.2})\Big(P^1|R^{1.2.2}, e^1 g^{1.2.2} \displaystyle\prod_{n\in\NN}(n.1.2.2\leftrightarrow n.1.1)\Big)\equiv_\alpha (X^{2.1}, x^{2.1})$ for some $(X, x)\in B$, which implies that $(\nu_1)(Q|X^1, fx^1)\in C^2$ and applying $n\mapsto n.1.2$ we get $(\nu_{1.1.2})(Q^{1.2}|X^{1.1.2}, f^{1.2} x^{1.1.2})\in C^{2.1.2}$. In particular, $(\nu_{1.1.2})(Q^{1.2}|X^{1.1.2}, f^{1.2} x^{1.1.2})=(Y^{2.1.2}, y^{2.1.2})$ for some $(Y, y)\in C$.

  Since $\NN^{2.1}$ is disjoint with $\NN^{1.1.2}\biguplus \NN^{1.2}$ we have \begin{equation*}
    \begin{split}
      \Big(Q^{1.2}|X^{1.1.2}, f^{1.2} x^{1.1.2}\Big)&\equiv_\alpha\\ 
      (\nu_{2.1})\Big(Q^{1.2}|X^{2.1}, f^{1.2}x^{2.1}\displaystyle\prod_{n\in\NN}(n.2.1\leftrightarrow n.1.1.2)\Big)& \equiv_\alpha\\
      (\nu_{2.1})\Big(\big(Q^{1.2}, f^{1.2}\displaystyle\prod_{n\in\NN}(n.2.1\leftrightarrow n.1.1.2)\big)\Big|(\nu_{1.1})(\nu_{1.2.2})\Big(P^1|R^{1.2.1}, e^1 g^{1.2.2} \displaystyle\prod_{n\in\NN}(n.1.2.2\leftrightarrow n.1.1)\Big)\Big)&\equiv_\alpha\\ (\nu_{2.1})(\nu_{1.1})(\nu_{1.2.2})\Big(Q^{1.2}|P^1|R^{1.2.1}, f^{1.2} e^1 g^{1.2.2} \displaystyle\prod_{n\in\NN}(n.1.2.2\leftrightarrow n.1.1)(n.2.1\leftrightarrow n.1.1.2)\Big)
    \end{split}
  \end{equation*}
  And then we deduce  $(\nu_{1.1.2})(\nu_{2.1})(\nu_{1.1})(\nu_{1.2.2})\Big(Q^{2.1}|P^1|R^{1.2.1}, f^{2.1} e^1 g^{1.2.2} \displaystyle\prod_{n\in\NN}(n.1.2.2\leftrightarrow n.1.1)(n.2.1\leftrightarrow n.1.1.2)\Big)\equiv_\alpha(Y^{2.1.2}, y^{2.1.2})\in C^{2.1.2}$. This implies that $(\nu_{2.2.2})(S^{2.2.2}|Y^{2.2.2}, h^{2.2.2} y^{2.2.2})\in\bbot$. Again reasoning as in {\bf Proposition} \ref{pr:nu-1-restriction-processes} we have $(\nu_{2.1.2})\Big(Y^{2.1.2}|S^{2.2.2}, y^{2.1.2} h^{2.2.2} \displaystyle\prod_{n\in\NN}(n.2.1.2\leftrightarrow n.2.2.2)\Big)\equiv_\alpha (Y^{2.2.2} S^{2.2.2}, y^{2.2.2} h^{2.2.2})$ and then  $(\nu_{2.2.2})(\nu_{2.1.2})\Big(Y^{2.1.2}|S^{2.2.2}, y^{2.1.2} h^{2.2.2} \displaystyle\prod_{n\in\NN}(n.2.1.2\leftrightarrow n.2.2.2)\Big)\equiv_\alpha (\nu_{2.2.2})(S^{2.2.2}|Y^{2.2.2}, h^{2.2.2} y^{2.2.2})\in\bbot$. 

  Finally, as before observe that: 
  \begin{equation*}
    \begin{split}
      (\nu_{2.2.2})(\nu_{2.1.2})\Big(Y^{2.1.2}|S^{2.2.2}, y^{2.1.2} h^{2.2.2} \displaystyle\prod_{n\in\NN}(n.2.1.2\leftrightarrow n.2.2.2)\Big)&\equiv_\alpha\\
      (\nu_{2.2.2})(\nu_{2.1.2})\Big[\big(S^{2.2.2}, h^{2.2.2}\displaystyle\prod_{n\in\NN}(n.2.1.2\leftrightarrow n.2.2.2)\big)\Big|&\\ (\nu_{1.1.2})(\nu_{2.1})(\nu_{1.1})(\nu_{1.2.2})\Big(Q^{2.1}|P^1|R^{1.2.1}, f^{2.1} e^1 g^{1.2.2} \displaystyle\prod_{n\in\NN}(n.1.2.2\leftrightarrow n.1.1)(n.2.1\leftrightarrow n.1.1.2)\Big)\Big]&\equiv_\alpha\\
      (\nu_{2.2.2})(\nu_{2.1.2})(\nu_{1.1.2})(\nu_{2.1})(\nu_{1.1})(\nu_{1.2.2})\Big(Q^{2.1}|P^1|R^{1.2.1}|S^{2.2.2}, f^{2.1} e^1 g^{1.2.2}h^{2.2.2}  \sigma_\tau\Big)
    \end{split}
  \end{equation*}
  all which ends the proof. 
\end{proof}

\begin{lemma}\label{lm:Adequation-lemma-v1}
  The following behaviours are inhabited by PWF processes which are pure fusions: 
  \begin{enumerate}
    \item\label{it01:Adequation-lemma-v1} $\bigcap_{A\in\mathcal{B}} A\multimap A$.
    \item\label{it02:Adequation-lemma-v1} $\bigcap_{A\in\mathcal{B}} A\multimap ({\bf 1}\otimes A)$ and $\bigcap_{A\in\mathcal{B}}({\bf 1}\otimes A)\multimap A$.
    \item \label{it03:Adequation-lemma-v1}$\bigcap_{A, B\in\mathcal{B}} (A\otimes B)\multimap (B\otimes A)$.
    \item \label{it04:Adequation-lemma-v1} $\bigcap_{A, B, C\in\mathcal{B}} (A\multimap B)\multimap (B\multimap C)\multimap A\multimap C$.
    \item \label{it05:Adequation-lemma-v1}$\bigcap_{A, B, C\in\mathcal{B}} ((A\otimes B)\otimes C)\multimap (A\otimes (B\otimes C))$.
  \end{enumerate}
\end{lemma}
\begin{proof}
  The proof is a consequence of {\bf Lemma}s \ref{lm:identityt-adequacy} to \ref{lm:composition-linear-arrow}. Observe that the fusions we found are \emph{uniform} on $A, B$ and $C$, which ends the proof. 
\end{proof}   

\begin{lemma}\label{lm:contraposition-adequacy}
  There exists a fusion $\sigma_\tau$ for some substitution $\tau$ s.t. $\sigma_\tau\in (A\multimap B)\multimap (B^\perp\multimap A^\perp)$ for all $A, B\subseteq \mathbb{P}$. 
\end{lemma}
\begin{proof}
  Let us consider the substitution $\tau:\NN^{1.1}\uplus \NN^{2.1}\to \NN^{2.2}\uplus \NN^{1.2}$ s.t. $\tau(n.1.1):= n.2.2$ and $\tau(n.2.1):=n.1.2$. 

  Let us consider $(P, e)\in A$, $(Q, f)\in B^\perp$ and $(H, h)\in (A\bullet B^\perp)^\perp$. It suffices to prove that \[\overline{\nu} (H^1| P^{2.2}| Q^{1.2}, e^{2.2} f^{1.2} h^1\sigma_\tau)\in \bbot\] 
  \[
    \begin{array}{rclc}
      \overline{\nu} (H^1| P^{2.2}| Q^{1.2}, e^{2.2} f^{1.2} h^1\sigma_\tau)&\equiv_\alpha& (\nu_1)(\nu_2)(H^1|P^{2.2}| Q^{1.2}, e^{2.2} f^{1.2} h^1 \sigma_\tau)&\equiv_\alpha\\ 
      (\nu_1)(H^1| P^{1.1} | Q^{2.1}, h^1 e^{1.1} f^{2.1})&\equiv_\alpha & \overline{\nu} (H| P^1|Q^2, h e^1 f^2)&\\
    \end{array}   
  \]
  which ends the proof since $(H, h)\in (A\bullet B^\perp)^\perp$ and $(P^1|Q^2, e^1 f^2)\in (A\bullet B^\perp)$.
\end{proof}

\begin{lemma}\label{lm:context-adequacy}
  There exists a fusion $\sigma_\tau$ for some substitution $\tau$ s.t. $\sigma_\tau\in (A\multimap B)\multimap (A\otimes C\multimap B\otimes C)$ for all $A, B, C\subseteq \mathbb{P}$. 
\end{lemma}
\begin{proof}
  Take $\tau:\NN^{1.1.2}\uplus \NN^{2.1.2}\uplus \NN^{1.2.2}\to \NN^{1.1}\uplus \NN^{2.2.2}\uplus \NN^{2.1}$ given by $\tau(n.1.1.2):=n.1.1$, $\tau(n.2.1.2):=n.2.2.2$ and $\tau(n.1.2.2):=n.2.1$. The proof is similar as in {\bf Lemma} \ref{lm:composition-linear-arrow}. 
\end{proof}

\section{Algebraic Structures for Concurrent Realizability}

\subsection{Conjunctive Structures and Implicative Structures} 

\begin{remark}\label{rk:Complete-join-meet}
  Given a complete join-semilattice $\mathcal{A}:=\left(\mathbb{A}, \bigcurlyvee, \preccurlyeq\right)$, defining $\forall\ \mathfrak{B}\subseteq \mathbb{A}\quad \bigcurlywedge \mathfrak{B}:=\bigcurlywedge_{b\in\mathfrak{B}}b:= \bigcurlyvee_{x\in lb(\mathfrak{B})}x$ we get a structure of complete lattice on  $\mathcal{A}$ (where $lb(\mathfrak{B})$ is the set of lower bounds of $\mathfrak{B}$).

  Dually, given a complete meet-semilattice $\mathcal{A}:=\left(\mathbb{A}, \bigcurlywedge, \preccurlyeq\right)$, defining $\forall \mathfrak{B}\subseteq \mathbb{A}\quad \bigcurlyvee\mathfrak{B}:=\bigcurlyvee_{b\in\mathfrak{B}}b:=\bigcurlywedge_{x\in ub(\mathfrak{B})}x$ we get a structure of complete lattice on  $\mathcal{A}$ (where $ub(\mathfrak{B})$ is the set of upper bounds of $\mathfrak{B}$).
\end{remark}

\begin{definition}\label{df:Conjunctive-Structure}\label{df:CS} (Miquey \cite{Miquey-thesis}, \cite{Miquey-duality}) A conjunctive Structure (CS) is $\mathcal{C}=(\mathbb{C}, \otimes, (\ )^\perp, \preccurlyeq)$ s.t.:
  \begin{enumerate}
    \item $(\mathbb{C}, \bigcurlyvee, \preccurlyeq)$ is a complete join semilattice (whose join is  $\bigcurlyvee$). 
    \item $\otimes$ is a binary monotone operation of $\mathbb{C}$ and $( )^\perp$ is a unary antimonotone function on $\mathbb{C}$. 
    \item The join operation $\bigcurlyvee$ is distributive w.r.t $\otimes$, i.e.: Given $a\in\mathbb{C}$ and $\mathfrak{B}\subseteq\mathbb{C}$ we have:
      \begin{itemize}
        \item $\bigcurlyvee_{b\in\mathfrak{B}}(a\otimes b)=a\otimes \Big(\bigcurlyvee_{b\in \mathfrak{B}}b\Big)$
        \item $\bigcurlyvee_{b\in\mathfrak{B}}(b\otimes a)=\Big(\bigcurlyvee_{b\in \mathfrak{B}}b\Big)\otimes a$
      \end{itemize}
    \item The orthogonal map $( )^\perp$ satisfies the De Morgan law $\Big(\bigcurlyvee_{b\in\mathfrak{B}}b\Big)^\perp=\bigcurlywedge_{b\in\mathfrak{B}} b^\perp$\footnote{The meet exists by the {\bf Remark} \ref{rk:Complete-join-meet}}.
    \item The orthogonal map is involutive, i.e.: $a^{\perp\perp}=a$ for all $a\in\mathbb{C}$\footnote{On Miquey's thesis \cite{Miquey-thesis} and subsequent work \cite{Miquey-duality}, the orthogonal map is not involutive.}
  \end{enumerate}
\end{definition}

\begin{definition}
  Given a (CS) $\mathcal{C}:=(\mathbb{C}, \otimes, ()^\perp, \preccurlyeq)$, we define the following connectors and quantifiers on $\mathcal{C}$:
  \begin{align*}
    a\parr b&:= (a^\perp\otimes b^\perp)^\perp\\
    a\multimap b&:=(a\otimes b^\perp)^\perp%=a^\perp\parr b
    \\
    a*b&:=\bigcurlywedge \{c\in \mathbb{C}\ |\ a\preccurlyeq b\multimap c\}\\
    \exists F & := \bigcurlyvee_{a\in\mathbb{C}} F(a) \text{, where $F:\mathbb{C}\to\mathbb{C}$ is a function.}
  \end{align*}

  By {\bf Remark} \ref{rk:Complete-join-meet} $\mathcal{C}:=(\mathbb{C}, \otimes, ()^\perp, \preccurlyeq)$ is a complete lattice. We denote by ${\bf 0}$ its bottom element and as ${\bf T}$ its top element. 
\end{definition}

\begin{proposition}\label{pr:monotonies-parr-arrow-exists}
  Let us consider a (CS) $\mathcal{C}:=(\mathbb{C}, \otimes, ()^\perp, \preccurlyeq)$. Then: 
  \begin{enumerate}
    \item\label{pr:monotonies-parr-arrow-exists:it1} $\parr$ is monotone on both arguments. 
    \item\label{pr:monotonies-parr-arrow-exists:it2} $\multimap$ is monotone on the right side and anti-monotone on the left side.
    \item\label{pr:monotonies-parr-arrow-exists:it3} $\exists$ is monotone on its (functional) argument.
    \item\label{pr:monotonies-parr-arrow-exists:it4} $a\multimap b=a^\perp\parr b$. 
  \end{enumerate}
\end{proposition}
\begin{proof}
  Is straightforward. See \cite{Miquey-thesis}. 
\end{proof}

\begin{proposition}\label{pr:CCS-distributive-laws}
  Let us consider a (CS) $\mathcal{C}:=(\mathbb{C}, \otimes, ()^\perp, \preccurlyeq)$. Then
  \begin{enumerate}
    \item\label{pr:CCS-distributive-laws:it1} $\Big(\bigcurlywedge_{b\in\mathfrak{B}}b\Big)^\perp=\bigcurlyvee_{b\in\mathfrak{B}}b^\perp$.
    \item\label{pr:CCS-distributive-laws:it2} $a\multimap \bigcurlywedge_{b\in\mathfrak{B}}b = \bigcurlywedge_{b\in\mathfrak{B}} (a\multimap b)$.
    \item\label{pr:CCS-distributive-laws:it3} $\Big(\bigcurlyvee_{b\in\mathfrak{B}} b\Big)\multimap a=\bigcurlywedge_{b\in\mathfrak{B}} (b\multimap a)$\footnote{This is not true for the Miquey's (CS) because is a consequence of $()^\perp$ being involutive.}. 
  \end{enumerate}
\end{proposition}
\begin{proof}\ 
  \begin{enumerate}
    \item Since $\Big(\bigcurlyvee_{b\in\mathfrak{B}}b^\perp\Big)^\perp = \bigcurlywedge_{b\in\mathfrak{B}} b^{\perp\perp}=\bigcurlywedge_{b\in\mathfrak{B}} b$, taking $(\ )^\perp$ we get $\bigcurlyvee_{b\in\mathfrak{B}}(b^\perp)=\Big(\bigcurlywedge_{b\in\mathfrak{B}} b\Big)^\perp$.
    \item $a\multimap \bigcurlywedge_{b\in\mathfrak{B}}b := \Big(a\otimes \big(\bigcurlywedge_{b\in\mathfrak{B}}b\big)^\perp\Big)^\perp = \Big(a\otimes \bigcurlyvee_{b\in\mathfrak{B}}b^\perp\Big)^\perp = \Big(\bigcurlyvee_{b\in\mathfrak{B}}a\otimes b^\perp\Big)^\perp = \bigcurlywedge_{b\in\mathfrak{B}} (a\otimes b^\perp)^\perp =: \bigcurlywedge_{b\in\mathfrak{B}} (a\multimap b)$.
    \item $\Big(\bigcurlyvee_{b\in\mathfrak{B}} b\Big)\multimap a := \Big(\Big(\bigcurlyvee_{b\in\mathfrak{B}} b\Big)\otimes a^\perp\Big)^\perp=\Big(\bigcurlyvee_{b\in\mathfrak{B}} b\otimes a^\perp\Big)^\perp = \bigcurlywedge_{b\in\mathfrak{B}}(b\otimes a^\perp)^\perp=:\bigcurlywedge_{b\in\mathfrak{B}} b\multimap a$. 
  \end{enumerate}
\end{proof}

\begin{definition}\label{df:IS}
  (Miquel \cite{Miquel-IA}) An implicative structure (IS) is $\mathcal{C}:=(\mathbb{C}, \to, \preccurlyeq)$ s.t.:
  \begin{itemize}
    \item $(\mathbb{C}, \preccurlyeq)$ is a complete meet-semilattice.
    \item $\to$ is a binary operation that is monotone on the second argument and antimonotone on the first argument.
    \item The meet operation $\bigcurlywedge$ is distributive w.r.t. $\to$, i.e.: $a\to \bigcurlywedge_{b\in\mathfrak{B}}b=\bigcurlywedge_{b\in\mathfrak{B}} (a\to b)$ for all $a\in\mathbb{C}$ and $\mathfrak{B}\subseteq \mathbb{C}$. 
  \end{itemize}
  An implicative structure is \emph{compatible with joins} iff $\Big(\bigcurlyvee_{b\in\mathfrak{B}} b\Big)\to a=\bigcurlywedge_{b\in\mathfrak{B}} (b\to a)$ for all $a\in\mathbb{C}$ and $\mathfrak{B}\subseteq \mathbb{C}$.
\end{definition}

\begin{remark}\label{rk:CCS-IS}
  Let us consider $\mathcal{C}:=(\mathbb{C}, \otimes, ()^\perp, \preccurlyeq)$ a (CS). Then $(\mathbb{C}, \multimap, \preccurlyeq)$ (with $a\multimap b:=(a\otimes b^\perp)^\perp$) is an~(IS) compatible with joins.
\end{remark}
\begin{proof}
  It is a direct consequence of {\bf Proposition} \ref{pr:CCS-distributive-laws}.\ref{pr:CCS-distributive-laws:it1},  \ref{pr:CCS-distributive-laws}.\ref{pr:CCS-distributive-laws:it2} and \ref{pr:CCS-distributive-laws}.\ref{pr:CCS-distributive-laws:it3}.  
\end{proof}

\begin{proposition}\label{rk:unit-counit-adj}
  Let us consider a (CS) $\mathcal{C}=(\mathbb{C}, \otimes, ()^\perp, \preccurlyeq)$. Then: 
  \begin{enumerate}
    \item\label{rk:unit-counit-adj:it1} $*$ is monotone in both arguments. 
    \item\label{rk:unit-counit-adj:it2} $(a\multimap b)*a\preccurlyeq b$.
    \item\label{rk:unit-counit-adj:it3} $a\preccurlyeq b\multimap a*b$.
    \item\label{rk:unit-counit-adj:it4} $a*b:=\min\{c\in\mathbb{C}\ |\ a\preccurlyeq b\multimap c\}$.
    \item\label{rk:unit-counit-adj:it5} $a*b\preccurlyeq c$ iff $a\preccurlyeq b\multimap c$. 
  \end{enumerate}
\end{proposition} 
\begin{proof}
  All is a direct consequence of the fact that $(\mathbb{A}, \multimap, \preccurlyeq)$ is an (IS). The proof is in \cite{Miquel-IA}.  
\end{proof}

\begin{definition}
  Let us consider a (CS) $\mathcal{C}:=(\mathbb{C},\otimes,  ()^\perp, \preccurlyeq)$. Define the following Logical Combinators\footnote{The combinators $S_3, S_4, S_5$ for (CS)'s are defined by \'E. Miquey\cite{Miquey-duality}.}associated to~$\mathcal{C}$:
  \begin{itemize}
    \item $S_3:=\bigcurlywedge_{a,b\in\mathbb{C}} a\otimes b\multimap b\otimes a$.
    \item $S_4:=\bigcurlywedge_{a,b,c\in\mathbb{C}} (a\multimap b)\multimap (b\multimap c)\multimap a\multimap c$.
    \item $S_5:=\bigcurlywedge_{a,b,c\in\mathbb{C}} ((a\otimes b)\otimes c)\multimap (a\otimes (b\otimes c))$.
  \end{itemize}
  We will consider (CS)'s with a distinguished element ${\bf 1}\in\mathbb{C}$ called \emph{unit}. For a (CS) with unit $\mathcal{C}=(\mathbb{C}, \otimes, ()^\perp, {\bf 1}, \preccurlyeq)$ we define the combinators:
  \begin{itemize}
    \item $S_6:=\bigcurlywedge_{a\in\mathbb{C}} a\multimap ({\bf 1}\otimes a)$ 
    \item $S_7:=\bigcurlywedge_{a\in\mathbb{C}} ({\bf 1}\otimes a)\multimap a$. 
  \end{itemize}
  For $\mathcal{C}$ a (CS) we define the combinators of $\mathcal{C}$ as the set $\operatorname{Comb}(\mathcal{C}):=\{S_3, S_4, S_5, S_6, S_7\}$. 
\end{definition}

\begin{definition}\label{df:CCA}
  A \emph{Conjunctive Algebra} (CA) is a couple $(\mathcal{C}, S)$ where: $\mathcal{C}=(\mathbb{C}, \otimes, |, ()^\perp, {\bf 1}, \preccurlyeq)$ is a (CPS), $S\subseteq \mathbb{C}$ and satisfies the following rules:
  \begin{center}
    \begin{tabular}{rcr}
      \begin{prooftree}
        c\in\operatorname{Comb}(\mathcal{C})
        \justifies
        c\in\mathcal{S}
        \using(\text{ax})
      \end{prooftree}&&
      \begin{prooftree}
        a\preccurlyeq b\quad a\in\mathcal{S}
        \justifies
        b\in\mathcal{S}
        \using(\text{upc})
      \end{prooftree}\\
          &&\\
          \begin{prooftree}
            a\multimap b\in\mathcal{S}\quad
            a\in\mathcal{S}
            \justifies
            b\in\mathcal{S}
            \using(\text{mp})
          \end{prooftree}&&
          \begin{prooftree}
            a\multimap b\in\mathcal{S}
            \justifies
            a\otimes c\multimap b\otimes c\in\mathcal{S}
            \using(\text{ctx})
          \end{prooftree}
          \\
          &&\\
          \begin{prooftree}
            a\multimap b\in\mathcal{S}
            \justifies
            b^\perp\multimap a^\perp\in\mathcal{S}
            \using(\text{ctr})
          \end{prooftree}&&
          \begin{prooftree} 
            \justifies 
            {\bf 1}\in\mathcal{S} 
            \using(\text{unit}) 
          \end{prooftree}
          \\
    \end{tabular}
  \end{center}
\end{definition}

\begin{remark}
  The definition of (CA)'s we give here is \emph{mutatis mutandis} the one given by \'E. Miquey on his PhD thesis \cite{Miquey-thesis} but adapted to the context of a linear calculus. The definition of \emph{conjunctive algebra} (CA) given by Miquey is a couple $(\mathcal{A}, \mathcal{S})$ where $\mathcal{A}$ is a conjunctive structure, $\mathcal{S}\subseteq \mathbb{A}$ contains the (linear) combinators $S_3, S_4$ and $S_5$, satisfies the rules (upc) and (mp) (respectively: \emph{upper closure} and  \emph{modus ponens closure}) and contains the (non-linear) combinators for weakening and contraction:
  \begin{itemize}
    \item $S_1:=\bigcurlywedge_{a\in\mathbb{A}} a\multimap (a\otimes a)$. 
    \item $S_2:=\bigcurlywedge_{a,b\in\mathbb{A}} (a\otimes b)\multimap a$. 
  \end{itemize}
  We do not use $S_1$ and $S_2$ in order to get a linear type system according to concurrency. In his thesis, Miquey also proves that combinators $S_6, S_7$ being to the separator, but his proof lies in the use of the combinators $S_1, S_2$. Similarly, the rules (ctx) and (ctr) are provable by by means of $S_1, S_2$. Here we explicitly add them to the separator. 
\end{remark}

\begin{proposition}\label{pr:*-S-closure}
  Let us consider a (CA) $(\mathcal{C}, \mathcal{S})$. If $a, b\in\mathcal{S}$, then $a*b\in\mathcal{S}$. 
\end{proposition}
\begin{proof} By {\bf Remark}  \ref{rk:unit-counit-adj}.\ref{rk:unit-counit-adj:it2} $a\preccurlyeq b\multimap a*b$. Thus we can derive: 
  \begin{center}
    \begin{prooftree}
      \[
        a\in\mathcal{S}\quad
        a\preccurlyeq b\multimap a*b
        \justifies
        b\multimap a*b\in\mathcal{S}
      \using(\text{upc})\]
      \quad
      b\in\mathcal{S}
      \justifies
      a*b\in\mathcal{S}
      \using(\text{mp})
    \end{prooftree}
  \end{center}

  In fact, (mp) is equivalent to the closure under $*$ (in presence of (upc)): By {\bf Remark} \ref{rk:unit-counit-adj}.\ref{rk:unit-counit-adj:it1} we get $(a\multimap b)*a\preccurlyeq b$. Then, using the closure of $\mathcal{S}$ under $*$ (app) we can derive: 
  \begin{center}
    \begin{prooftree}
      \[
        a\multimap b\in\mathcal{S}
        \quad
        a\in\mathcal{S}
        \justifies
        (a\multimap b)*a\in\mathcal{S}
        \using(\text{app})
      \]
      \quad
      (a\multimap b)*a\preccurlyeq b
      \justifies
      b\in\mathcal{S}
      \using(\text{upc})
    \end{prooftree}
  \end{center}
\end{proof}

\begin{definition}\label{df:composition} 
  Let us consider a (CA) $(\mathcal{C}:=(\mathbb{C}, \otimes, ()^\perp, {\bf 1},  \preccurlyeq), \mathcal{S})$. We define:
  \[
    \begin{array}{rcl}
      \forall a,b\in\mathbb{C}\quad \Hom(a,b)&:=&\{s\in\mathcal{S}\ |\ s\preccurlyeq a\multimap b\}\\
      \forall s, t\quad s\in\Hom(a,b),\quad  t\in\Hom(b,c)\quad \Rightarrow\quad t\circ s&:=&S_4*s*t\\
    \end{array}
  \]
\end{definition}
\begin{remark}\label{rk:separator-closed-under-composition}
  Since $\mathcal{S}$ is closed under $*$, we have: $\forall s,t\quad s\in\Hom(a,b), \quad t\in\Hom(b,c)\quad \Rightarrow\quad  t\circ s\in\Hom(a,c)$. In particular, $\circ:\Hom(a,b)\times \Hom(b,c)\to \Hom(a,c)$. 
\end{remark} 

\begin{proposition}\label{pr:parr-orthogonal-rules}
  Let us consider a (CA) $(\mathcal{C}, \mathcal{S})$, where $\mathcal{C}=(\mathbb{C}, \otimes, |, ()^\perp, {\bf 1}, \preccurlyeq)$. Then following elements belong to $\mathcal{S}$ for all $a,b,c\in\mathbb{C}$:  
  \begin{enumerate}
    \item\label{pr:parr-orthogonal-rules:it1} $(a\otimes (b\otimes c))\multimap ((a\otimes b)\otimes c)$.
    \item\label{pr:parr-orthogonal-rules:it2} $(a\otimes b)^\perp\multimap (b\otimes a)^\perp$.
    \item\label{pr:parr-orthogonal-rules:it3} $(a\otimes(b\otimes c))^\perp\multimap ((a\otimes b)\otimes c)^\perp$ and $((a\otimes b)\otimes c)^\perp\multimap (a\otimes (b\otimes c))^\perp$.
    \item $a\multimap b\multimap (a\otimes b)$.
    \item\label{pr:parr-orthogonal-rules:it4} $(a\parr b)\multimap (b\parr a)$.
    \item\label{pr:parr-orthogonal-rules:it5} $(a\parr b)\parr c\multimap a\parr (b\parr c)$ and $a\parr (b\parr c)\multimap (a\parr b)\parr c$.
    \item\label{pr:parr-orthogonal-rules:it6} $(a\parr {\bf 1}^\perp)\multimap a$ and $a\multimap (a\parr {\bf 1}^\perp)$.
  \end{enumerate}
\end{proposition}
\begin{proof}\ 
  \begin{enumerate}
    \item\

      %\begin{center}
      \begin{tabular}{rrcll}
        By $S_3$&$a\otimes (b\otimes c)$ &$\multimap$ & $(b\otimes c)\otimes a$& $\in\mathcal{S}$\\
        By $S_5$&$(b\otimes c)\otimes a$ &$\multimap$ & $b\otimes (c\otimes a)$ & $\in\mathcal{S}$\\
        By $S_3$ & $b\otimes (c\otimes a)$ & $\multimap$ & $(c\otimes a)\otimes b$& $\in\mathcal{S}$\\
        By $S_5$ & $(c\otimes a)\otimes b$ & $\multimap$ & $c\otimes (a\otimes b)$ & $\in\mathcal{S}$\\
        By $S_3$ & $c\otimes (a\otimes b)$ & $\multimap$ & $(a\otimes b)\otimes c$ & $\in\mathcal{S}$\\
      \end{tabular}
      %\end{center}

      We conclude by {\bf Remark} \ref{rk:separator-closed-under-composition} that $a\otimes (b\otimes c)\multimap (a\otimes b)\otimes c\in\mathcal{S}$. 
    \item\ \\
      \begin{prooftree}
        \[
          \[
            \justifies
            S_3\in\mathcal{S}
            \using(\text{ax})
          \]
          \quad
          S_3\preccurlyeq (b\otimes a)\multimap (a\otimes b)
          \justifies
          (b\otimes a)\multimap (a\otimes b)\in\mathcal{S}
          \using(\text{upc})
        \]
        \justifies
        (a\otimes b)^\perp\multimap (b\otimes a)^\perp\in\mathcal{S}
        \using(\text{ctr})
      \end{prooftree}
    \item\ \\
      \begin{prooftree}
        \[
          \[
            \justifies
            S_5\in\mathcal{S}
            \using(\text{ax})
          \]
          \quad 
          S_5\preccurlyeq ((a\otimes b)\otimes c)\multimap (a\otimes (b\otimes c))
          \justifies
          ((a\otimes b)\otimes c)\multimap (a\otimes (b\otimes c))\in\mathcal{S}
          \using(\text{upc})
        \]
        \justifies
        (a\otimes (b\otimes c))^\perp\multimap ((a\otimes b)\otimes c)^\perp\in\mathcal{S}
        \using(\text{ctr})
      \end{prooftree}
      The other sentence result of \ref{pr:parr-orthogonal-rules:it1}. 
    \item\ 

      \begin{tabular}{lrcll} 
        By \ref{pr:parr-orthogonal-rules:it1}&  $a\otimes (b\otimes (a\otimes b)^\perp)$&$\multimap$&$(a\otimes b)\otimes (a\otimes b)^\perp$&$\in \mathcal{S}$\\  
        By \ref{pr:parr-orthogonal-rules:it2}&  $((a\otimes b)\otimes (a\otimes b)^\perp))^\perp$&$\multimap$&$(a\otimes (b\otimes (a\otimes b)^\perp)^\perp$&$\in \mathcal{S}$\\  
      \end{tabular}  
      Since $((a\otimes b)\otimes (a\otimes b)^\perp))^\perp=(a\otimes b)\multimap (a\otimes b)\in\mathcal{S}$, by the (mp)-rule we get $((a\otimes b)\otimes (a\otimes b)^\perp))^\perp=a\multimap b\multimap (a\otimes b)\in\mathcal{S}$.
    \item Is a particular case of \ref{pr:parr-orthogonal-rules:it2}.
    \item Is a particular case of \ref{pr:parr-orthogonal-rules:it3}.
    \item \

      \begin{prooftree}
        \[
          \[
            \justifies
            S_6\in\mathcal{S}
            \using(\text{ax})
          \]
          \quad
          S_6\preccurlyeq a^\perp\multimap (a^\perp\otimes {\bf 1})
          \justifies
          a^\perp\multimap (a^\perp\otimes {\bf 1})\in\mathcal{S}
          \using(\text{upc})
        \]
        \justifies
        (a\parr {\bf 1}^\perp)\multimap a 
        \using(\text{ctr})
      \end{prooftree} The converse way has a similar proof.
  \end{enumerate}   
\end{proof}

\begin{proposition}\label{pr:semi-dist}
  Let us consider a (CA) $(\mathcal{C}, \mathcal{S})$, where $\mathcal{C}=(\mathbb{C}, \otimes, ()^\perp, {\bf 1}, \preccurlyeq)$. 

  Then 
  $\forall a,b,c\in\mathbb{C}\quad (a\parr b)\otimes c\multimap a\parr (b\otimes c)\in\mathcal{S}$
\end{proposition}
\begin{proof}
  By definition, we must prove $\Big(\big((a^\perp \otimes b^\perp)^\perp\otimes c\big)\otimes \big(a^\perp\otimes(b\otimes c)^\perp\big)\Big)^\perp\in\mathcal{S}$.

  \begin{tabular}{rr@{\;}c@{\;}ll}
    By $S_5$ & $\big((a^\perp \otimes b^\perp)^\perp\otimes c\big)\otimes\big(a^\perp\otimes (b\otimes c)^\perp\big)$&$\multimap$ & $(a^\perp \otimes b^\perp)^\perp\otimes\big(c\otimes\big(a^\perp\otimes (b\otimes c)^\perp\big)\big)$& $\in\mathcal{S}$\\
    By \ref{pr:parr-orthogonal-rules}.\ref{pr:parr-orthogonal-rules:it1}+(ctx)& $(a^\perp \otimes b^\perp)^\perp\otimes\big(c\otimes\big(a^\perp\otimes (b\otimes c)^\perp\big)\big)$ & $\multimap$& $(a^\perp \otimes b^\perp)^\perp\otimes\big((c\otimes a^\perp)\otimes (b\otimes c)^\perp\big)$&$\in\mathcal{S}$\\
    By $S_3$+(ctx) & $(a^\perp \otimes b^\perp)^\perp\otimes\big((c\otimes a^\perp)\otimes (b\otimes c)^\perp\big)$ & $\multimap$ & $(a^\perp \otimes b^\perp)^\perp\otimes\big((b\otimes c)^\perp\otimes (a^\perp\otimes c)\big)$&$\in\mathcal{S}$\\
    By \ref{rk:separator-closed-under-composition} &
    $\big((a^\perp \otimes b^\perp)^\perp\otimes c\big)\otimes\big(a^\perp\otimes (b\otimes c)^\perp\big)$&$\multimap$ &
    $(a^\perp \otimes b^\perp)^\perp\otimes\big((b\otimes c)^\perp\otimes (a^\perp\otimes c)\big)$&$\in\mathcal{S}$\\
    By (ctr) & $\Big((a^\perp \otimes b^\perp)^\perp\otimes\big((b\otimes c)^\perp\otimes (a^\perp\otimes c)\big)\Big) ^\perp$&$\multimap$&$\Big(\big((a^\perp \otimes b^\perp)^\perp\otimes c\big)\otimes\big(a^\perp\otimes (b\otimes c)^\perp\big)\Big)^\perp$&$\in\mathcal{S}$\\
  \end{tabular}

  By definition, it means that $\Big((a^\perp\multimap b)\multimap \big((b\multimap c^\perp)\multimap (a^\perp\multimap c^\perp)\big)\Big)\quad \multimap\quad \Big((a\parr b)\otimes c \multimap a\parr(b\otimes c)\Big)$.

  Since $S_4\preccurlyeq \Big((a^\perp\multimap b)\multimap \big((b\multimap c^\perp)\multimap (a^\perp\multimap c^\perp)\big)\Big)$, applying (mp) we get: $(a\parr b)\otimes c \multimap a\parr(b\otimes c)\in\mathcal{S}$.  
\end{proof}

\begin{proposition}\label{pr:soundness}
  Let us consider a (CA) $(\mathcal{C},\mathcal{S})$. Then:
  \begin{enumerate}
    \item $\forall a\in\mathbb{C}\quad a\multimap a\in\mathcal{S}$.
    \item $\forall a_1, \dots, a_k\in\mathbb{C}\quad (a_1\parr \dots \parr a_k)\multimap (a_{\sigma(1)}\parr \dots \parr a_{\sigma(k)})\in\mathcal{S}$ for all $\sigma$ permutation of $k$ elements.  
    \item $\forall a, b, g\in\mathbb{C}\quad g\parr a\preccurlyeq g\parr (a\curlyvee b)$ and hence $(g\parr a)\multimap (g\parr (a\curlyvee b))\in\mathcal{S}$. 
    \item $\forall a,b,g,d\in\mathbb{C}\quad (g\parr a)\otimes (b\parr d) \multimap g\parr ((a\otimes b)\parr d)\in\mathcal{S}$ and $(g\parr a)\multimap (b\parr d)\multimap g\parr((a\otimes b)\parr d)\in \mathcal{S}$.
    \item $\forall a,g,d\in\mathbb{C}\quad (g\parr a)\multimap (a^\perp \parr d)\multimap (g\parr d)\in\mathcal{S}$.
    \item ${\bf 1}\in\mathcal{S}$. 
    \item $\forall g, b\in\mathbb{C}\quad \forall F\in\mathbb{C}^\mathbb{C}\quad g\parr F(b)\preccurlyeq g\parr \exists F$.
  \end{enumerate}
\end{proposition}
\begin{proof}\ 
  \begin{enumerate}
    \item\

      \begin{tabular}{rcccl}
        By $S_6$ & $a$&$\multimap$& $a\otimes {\bf 1}$&$\in\mathcal{S}$\\
        By $S_7$ & $a\otimes {\bf 1}$ & $\multimap$ &$a$ & $\in\mathcal{S}$\\
        By \ref{rk:separator-closed-under-composition} & $a$ & $\multimap$ & $a$ & $\in\mathcal{S}$
      \end{tabular}
    \item Is a consequence of $S_3, S_4, S_5\in\mathcal{S}$, the {\bf Proposition} \ref{pr:parr-orthogonal-rules}. itemii \ref{pr:parr-orthogonal-rules:it4} and  \ref{pr:parr-orthogonal-rules:it5} and {\bf Remark} \ref{rk:separator-closed-under-composition}.
    \item By {\bf Proposition} \ref{pr:monotonies-parr-arrow-exists}.\ref{pr:monotonies-parr-arrow-exists:it1}.  
    \item \

      \begin{tabular}{rrcll}
        By \ref{pr:semi-dist} & $(g\parr a)\otimes (b\parr d)$&$\multimap$ & $g\parr (a\otimes (b\parr d))$& $\in\mathcal{S}$\\
        By $S_3$+\ref{pr:semi-dist} & $g\parr (a\otimes (b\parr d))$& $\multimap$ & $g\parr ((a\otimes b) \parr d)$ & $\in\mathcal{S}$\\
        By \ref{rk:separator-closed-under-composition} & $(g\parr a)\otimes (b\parr d)$&$\multimap$ & $g\parr ((a\otimes b) \parr d)$ & $\in\mathcal{S}$\\
      \end{tabular}

      Applying {\bf Lemma} \ref{pr:parr-orthogonal-rules}.\ref{pr:parr-orthogonal-rules:it4} and $S_4$ we get that $(g\parr a)\multimap (b\parr d)\multimap g\parr ((a\otimes b)\parr d)$. 
    \item By $S_4\preccurlyeq (g^\perp\multimap a)\multimap (a\multimap d) \multimap (g^\perp \multimap d)$ which, by definition of $\multimap$ and (upc), means that $(g\parr a)\multimap (a^\perp \parr d)\multimap (g\parr d)\in\mathcal{S}$.
    \item By {\bf Definition} \ref{df:CCA}, (unit)-rule. 
    \item By {\bf Proposition} \ref{pr:monotonies-parr-arrow-exists}.\ref{pr:monotonies-parr-arrow-exists:it3}. 
  \end{enumerate}
\end{proof}

\subsection{The types} 

\begin{definition}\label{df:MLL-rules}
  The language of multiplicative linear logic (MLL) is given by the following grammar:
  \[
    A, B ::= \one\ |\ X \ |\ A^\perp \ |\ A\otimes B \ |\ A\curlyvee B\ |\ \exists X.A 
  \]
  where $X$ is a variable of formul\ae.

  A (MLL) sequent is an expression of the shape $\vdash A_1, \dots, A_k$ where $A_1, \dots, A_k$ are formul\ae. We will abbreviate sequents by capital greek letters $\Gamma, \Delta, \Sigma, \dots$. We define the following proof-rules:
  \begin{center}
    \begin{tabular}{ll}
      \begin{prooftree}
        \justifies
        \vdash A^\perp, A
        \using(\text{ax})
      \end{prooftree}&
      \begin{prooftree}
        \vdash A_1, \dots, A_k
        \justifies
        \vdash A_{\sigma(1)}, \dots, A_{\sigma(k)}
        \using(\text{ex})
      \end{prooftree} where $\sigma$ is a permutation of $k$ elements\\
          &\\
          \begin{prooftree}
            \vdash \Gamma, A
            \justifies
            \vdash \Gamma, A\curlyvee B
            \using(\text{sub})
          \end{prooftree}&
          \begin{prooftree}
            \vdash\Gamma, A
            \quad
            \vdash A^\perp, \Delta
            \justifies
            \vdash \Gamma, \Delta
            \using(\text{cut}) 
          \end{prooftree}
          \\
          &\\
          \begin{prooftree}
            \justifies
            \vdash \one
            \using(\one)
          \end{prooftree}
          &
          \begin{prooftree}
            \vdash \Gamma, A
            \quad \vdash B, \Delta
            \justifies
            \vdash \Gamma, A\otimes B, \Delta
            \using(\otimes)
          \end{prooftree}\\
          &\\
          \begin{prooftree}
            \vdash \Gamma, A\{X:= B\}
            \justifies
            \vdash \Gamma, \exists X.A
            \using (\exists)
          \end{prooftree}
    \end{tabular}
  \end{center}
\end{definition}

\begin{definition}
  Let us consider a (CA) $(\mathcal{C}, \mathcal{S})$. An  \emph{assignment} $\mathcal{Y}$ is a function s.t. for each variable of formul\ae~$X$, its assignment $\mathcal{Y}(X)\in\mathbb{A}$. Given an   assingnment $\mathcal{Y}$, it define an \emph{interpretation} for the language of (MLL) as follows:

  \begin{tabular}{rcl}
    $\llbracket \one\rrbracket$ & $:=$ & $\one$\\
    $\llbracket X\rrbracket$ & $:=$ & $\mathcal{Y}(X)$\\
    $\llbracket A^\perp\rrbracket$ & $:=$ & $\llbracket A\rrbracket ^\perp$\\
    $\llbracket A\otimes B\rrbracket$ & $:=$ & $\llbracket A\rrbracket \otimes \llbracket B\rrbracket$\\
    $\llbracket A\curlyvee B\rrbracket$ & $:=$ & $\llbracket A\rrbracket \curlyvee \llbracket B\rrbracket$\\
    $\llbracket \exists X.A\rrbracket$ & $:=$ & $\exists \Big(B\mapsto \llbracket A\rrbracket\{X:=\llbracket B\rrbracket\}\Big)$\\
  \end{tabular}

  Given a sequent $\vdash A_1, \dots, A_k$, the interpretation $\llbracket A_1, \dots, A_k\rrbracket$ is defined as $\llbracket A_1\rrbracket$ if $k=1$ and $\llbracket A_1\rrbracket \parr \llbracket A_2, \dots, A_k\rrbracket$ otherwise. 
\end{definition}

\begin{remark}
  As usually, the interpretation $\llbracket A\rrbracket$ of a formula $A$ depends only upon the assignement of the free variables of $A$. The same is valid for sequents. 
\end{remark}

\begin{definition}\label{df:soundness-MLL}
  A proof-rule
  $\begin{prooftree} 
    \vdash \Sigma_1, \dots, \vdash\Sigma_h
    \justifies
    \vdash \Gamma
  \end{prooftree}$ is correct (or sound) w.r.t a (CPA) $(\mathcal{A}, \mathcal{S})$ together with an assignment $\mathcal{Y}$ iff $\llbracket \Gamma\rrbracket\in\mathcal{S}$ whenever $\llbracket \Sigma_1\rrbracket, \dots, \llbracket \Sigma_h\rrbracket\in\mathcal{S}$
\end{definition}

\begin{theorem}\label{th:soundness-MLL}
  The proof rules of (MLL) are correct for all (CPA). 
\end{theorem}
\begin{proof}
  The proof corresponds one by one with the sentences of {\bf Proposition} \ref{pr:soundness}. 
\end{proof}

\subsection{The algebraic model induced by the (PWF)}

\begin{proposition}\label{pr:PWF-as-CA-model}
  Let us consider $\mathcal{B}:=(\mathbb{B}, \otimes, ()^\perp, {\bf 1}, \subseteq)$ according with {\bf Definition}s  \ref{df:poles-orthogonality}, \ref{df:poles-orthogonality}, {\bf Proposition} \ref{pr:lattices}, {\bf Definition} \ref{df:operations-with-truth-values}. Take $\mathcal{S}:=\mathbb{B}\setminus\{\emptyset\}$. Then $(\mathcal{B}, \mathcal{S})$ is a (CA).
\end{proposition}
\begin{proof} We have already proven all we need. Let us recap the proof, step by step:  
  \begin{itemize}
    \item $(\mathbb{B}, \vee, \subseteq)$ is a complete join semilattice by {\bf Proposition} \ref{pr:lattices}.
    \item $\otimes$ is monotone on both arguments since $A\otimes B=(A\bullet B)^{\perp\perp}$, $\bullet$ is monotone by definition and $()^\perp$ is anti-monotone by {\bf Proposition} \ref{pr:poles-orthogonality}.\ref{it01:poles-orthogonality}. 
    \item $A\otimes \bigvee\mathfrak{B}=\bigvee_{B\in\mathfrak{B}} (A\otimes B)$ for all $\mathfrak{B}\subseteq \mathbb{B}$ and $A\in\mathbb{B}$ by  {\bf Proposition} \ref{lm:otimes-vee-compatibility}.
    \item $\left(\bigvee \mathfrak{B}\right)^\perp = \bigwedge_{B\in\mathfrak{B}}B^\perp$ for all $\mathfrak{B}\subseteq\mathbb{B}$ by {\bf Proposition}  \ref{pr:poles-orthogonality}.\ref{it03:poles-orthogonality} 
    \item Since $\mathbb{B}$ are the subsets of (PWF) closed under $()^{\perp\perp}$, we get by definition $A^{\perp\perp}=A$ for all $A\in\mathbb{B}$. 
  \end{itemize}
  So far we have proven that $\mathcal{B}:=(\mathbb{B}, \otimes, ()^\perp, {\bf 1}, \subseteq)$ is a (CS) with unit. To prove that $\mathcal{S}$ is a separator we use what we call the Adequation Lemma:
  \begin{itemize}
    \item All the combinators $S_3$ to $S_7$ are inhabited by {\bf Lemma}  \ref{lm:Adequation-lemma-v1} from item \ref{it02:Adequation-lemma-v1} to item \ref{it05:Adequation-lemma-v1} and then $S_3$ to $S_7$ belong $\mathcal{S}$. Then $\mathcal{S}$ satisfies the (ax)-rule on {\bf Definition} \ref{df:CCA}. Similarly, since $\one\in{\bf 1}$, then ${\bf 1}\in\mathcal{S}$ and then $\mathcal{S}$ satisfies the (unit)-rule. 
    \item $\mathcal{S}$ satisfies the (upc)-rule since $A\subseteq B$ where $A\neq\emptyset$ implies $B\neq\emptyset$.   
    \item We observe that there is a general mechanism to prove that $\mathcal{S}$ satisfies a rule of the shape $\begin{prooftree}
        G_1\in\mathcal{S}\quad \dots \quad G_k\in\mathcal{S}
        \justifies
        D\in\mathcal{S}
      \end{prooftree}$. Suppose we prove that $(P,e)\in G_1\multimap \dots \multimap G_k\multimap D\in\mathcal{S}$ for some (PWF)-process $(P,e)$ and that there are $(Q_1, f_1)\in G_1, \dots, (Q_k, f_k)\in G_k$. 

      On the other hand since $G_1\multimap \dots \multimap G_k\multimap D\subseteq G_1\multimap \dots \multimap G_k\multimap D$, by the adjoint $*_1$, we get
      \[
        \left(G_1\multimap \dots \multimap G_k\multimap D\right) *_1 G_1 *_1\dots *_1 G_k\subseteq D
      \]
      By this remark, we have that $(P, e)*_1 (Q_1, f_1)*_1\dots *_k (Q_k, f_k)\in D$ and then $D\in\mathcal{S}$.

      Applying this general mechanism we get:
      \begin{itemize}
        \item By {\bf Lemma} \ref{lm:Adequation-lemma-v1}.\ref{it01:Adequation-lemma-v1}, $I\in (A\multimap B)\multimap A\multimap B$ and then $\mathcal{S}$ satisfies the (mp)-rule.
        \item By {\bf Lemma} \ref{lm:context-adequacy} there is a $\sigma_\tau\in (A\multimap B)\multimap (A\otimes C)\multimap (B\multimap C)$ and then $\mathcal{S}$ satisfies the (ctx)-rule.
        \item By {\bf Lemma} \ref{lm:contraposition-adequacy} there is a $\sigma_\tau \in (A\multimap B)\multimap B^\perp\multimap A^\perp$ and  then $\mathcal{S}$ satisfies the (ctr)-rule. 
      \end{itemize}
  \end{itemize}        
\end{proof}

\subsection{Adding parallelism and combinators for pi-calculus}

\begin{definition}\label{df:CCS}
  A \emph{Conjunctive Parallel Structure} (CPS) is a Conjunctive Structure with unit $(\mathbb{C}, \otimes, ()^\perp, {\bf 1}, \preccurlyeq)$ together with a binary operation $(|):\mathbb{C}\times\mathbb{C}\to \mathbb{C}$ called  {\emph parallel composition} s.t. $(\mathbb{C}, |, {\bf 1})$ is  an abelian monoid, i.e.: 
  \begin{center}
    \begin{tabular}{rclcl}
      $(p|q)|r$ & $=$ & $p|(q|r)$ & & \\
      $p|{\bf 1}$ & $=$ & ${\bf 1}|p$ & $=$ & $p$\\
      $p|q$ & $=$ & $q|p$ & & \\
    \end{tabular}
  \end{center}
  And the condition of compatibility between $|$ and $\bigcurlyvee$ given by\footnote{This axiom is needed to define the right adjoint of parallel composition. At this point we don't know whether this axiom is independent from the rest or not, but it is satisfied by the model of behaviours defined from (PWF)-processes.}:
  \[
    \left(\bigcurlyvee \mathcal{B}\right)\Big| a\preccurlyeq \bigcurlyvee \left\{ b|a\ \ |\ b\in\mathcal{B}\right\}
  \]
  for all $\mathcal{B}\subseteq \mathbb{C}$.

  A (CPA) is a (CPS) $\mathcal{C}=(\mathbb{C}, \otimes, ()^\perp, |, {\bf 1},)$  together with a separator $\mathcal{S}$ s.t. $(\mathcal{C}, \mathcal{S})$ is a (CA). 
\end{definition}

\begin{definition}\label{df:rhd}
  Given a (CPA) $\mathcal{C}=(\mathbb{C}, \otimes, |, ()^\perp, \one, \preccurlyeq)$ we define $\rhd:\mathbb{C}\times\mathbb{C}\to\mathbb{C}$ by means of \[b\rhd c:=\bigcurlyvee \{x\in\mathbb{C}\ |\ x|b\preccurlyeq c\}\] 
\end{definition}

\begin{proposition}\label{pr:rhd-as-adjunction}
  Let us consider a (CPA) $\mathcal{C}=(\mathbb{C}, \otimes, |, ()^\perp, {\bf 1}, \preccurlyeq)$. Then for all $a,b,c\in\mathbb{C}\quad a|b\preccurlyeq c\Longleftrightarrow a\preccurlyeq b\rhd c$. 
\end{proposition}
\begin{proof}\ 
  \begin{itemize}
    \item[$(\Longrightarrow)$] $a|b\preccurlyeq c$ implies that $a\in\{x\in\mathbb{C} \ |\ x|b\preccurlyeq c\}$ and then $a\preccurlyeq b\rhd c$. 
    \item[$(\Longleftarrow)$] Since $|$ is monotone (because $*$ is monotone) we have $a|b\preccurlyeq (b\rhd c)|b=\left(\bigcurlyvee \{x\in\mathbb{C}\ |\ x|a\preccurlyeq c\}\right)|b\preccurlyeq\\ \bigcurlyvee \{x|b\ |\ x\in\mathbb{C}, x|b\preccurlyeq c\}\preccurlyeq c$, this using the compatibility between $|$ and $\bigcurlyvee$.  
  \end{itemize}
\end{proof}

\begin{proposition}\label{pr:PWF-as-CPA-model}
  The (CA) $(\mathbb{B}, \otimes,  ()^\perp, {\bf 1}, \subseteq)$ with separator $\mathcal{S}:=\mathbb{B}\setminus\{\emptyset\}$ (c.f.: {\bf Proposition} \ref{pr:PWF-as-CA-model}) is a (CPA) with the operator $\|$ (c.f.: {\bf Definition} \ref{df:operations-with-truth-values}).
\end{proposition}
\begin{proof}\ 
  \begin{itemize}
    \item Since $(\overline\Pi, |, \one)$ is an abelian monoid, then $(\mathbb{B}, \|, {\bf 1})$ is also. 
    \item $\left(\bigvee\mathfrak{B}\right) | A\subseteq \bigvee\left\{B\| A\ | B\in\mathfrak{B}\right\}$ for all $\mathfrak{B}\subseteq\mathbb{B}$ and $A\in\mathbb{B}$ by {\bf Lemma}  \ref{lm:compatibility-parallel-join} (observe that $B|A\subseteq B\| A$).
  \end{itemize}
\end{proof}

\begin{definition}\label{df:CCPS}
  A \emph{Combinatory Conjunctive Parallel Structure} (CCPS) is a (CPS) with unit $\mathcal{C}=(\mathbb{C}, \otimes, |, ()^\perp, {\bf 1}, \preccurlyeq)$ together with an injective function ${\bf M}:\NN\times \NN\to \mathbb{C}$. 
\end{definition}

\begin{definition}\label{df:Yoshida-Combi}
  Given a (CCPS)  $\mathcal{C}:=(\mathbb{C}, \otimes, |, ()^\perp, {\bf 1}, {\bf M}, \preccurlyeq)$ we define the \emph{Honda \& Yoshida Combinators} of $\mathcal{C}$ by means of:
  \begin{align*}
    K(a)&:=\bigcurlywedge_{x\in\NN}\big(M(a,x)\rhd {\bf 1}\big)\\
    F(a,b)&:=\bigcurlywedge_{x\in\NN}\big(M(a,x)\rhd M(b,x)\big)\\
    Bl(a,b)&:=\bigcurlywedge_{x\in\NN}\big(M(a,x)\rhd F(x,b)\big)\\
    Br(a,b)&:=\bigcurlywedge_{x\in\NN}\big(M(a,x)\rhd F(b,x)\big)\\
    D(a,b,c)&:=\bigcurlywedge_{x\in\NN}\big(M(a,x)\rhd M(b,x)|M(c,x)\big)\\
    S(a,b,c)&:=\bigcurlywedge_{x\in\NN}\big(M(a,x)\rhd F(b,c)\big)
  \end{align*}
\end{definition}

\begin{proposition}\label{pr:Yoshida-Comb-reductions}
  Given a (CCPS)  $\mathcal{C}=(\mathbb{C}, \otimes, |, ()^\perp, {\bf 1}, {\bf M}, \preccurlyeq)$ we have:
  \begin{align*}
    K(a)|M(a,x)&\preccurlyeq {\bf 1}\\
    F(a,b)|M(a,x)&\preccurlyeq M(b,x)\\
    Bl(a,b)|M(a,x)&\preccurlyeq F(x,b)\\
    Br(a,b)|M(a,x)&\preccurlyeq F(b,x)\\
    D(a,b,c)|M(a,x)&\preccurlyeq M(b,x)|M(c,x)\\
    S(a,b,c)|M(a,x)&\preccurlyeq F(b,c)
  \end{align*}
\end{proposition}

\begin{definition}\label{df:CCPA}
  A \emph{Combinatory Conjunctive Parallel Algebra} is a (CCPS) $\mathcal{C}=(\mathbb{C}, \otimes, |, ()^\perp, {\bf 1}, {\bf M}, \preccurlyeq)$ together with a \emph{separator} $\mathcal{S}\subseteq \mathbb{C}$ s.t. $((\mathbb{C}, \otimes, {\bf 1}, \preccurlyeq), \mathcal{S})$ is a (CPA) and all the Honda \& Yoshida combinators belong to $\mathcal{S}$. 
\end{definition}  
On secuential models of Classical Realizability, the set of realizers has associated an operational semantics given by reduction. This notion of reduction is correlated with the constructors/destructors for types: e.g.: the rules of introduction/elimination of implication corresponds respectively to abstraction and $\beta$-reduction on $\lambda$-calculus. In these models, to get the \emph{Adequation Lemma} we need to close the the poles by anti-reduction. 

So far we have not needed to introduce a notion of reduction because the constructors of Linear Logic here reflect the properties of connections and handling of names. However, to obtain a model of behaviours $\mathcal{B}$ where the Honda \& Yoshida's combinators defined by adjunction (c.f.: {\bf Definition} \ref{df:CCPA}) are inhabited, we ask the poles to be closed by anti-reduction. This choice follows the Beffara's design when defining the \emph{regular poles}  \cite{Beffara-thesis}. 
\begin{definition}\label{df:reduction} The reduction on (PWF) is defined as the least binary relation $\reduce$ s.t.: 
\begin{itemize}
    \item $\left(u^\uparrow(\vec{x}).P|v^\downarrow (\vec{x}).Q, e\right)\reduce \left((\nu \vec{x})(P|Q),e\right)$, whenever $u\simf{e} x$.
    \item Respects the constructions $|$ and $\nu$ of (PWF) (c.f.: {\bf Definition} \ref{df:constructors-on-PWF}).
    \item Is closed under $\equiv_\alpha$. 
\end{itemize}
\end{definition}
\begin{definition}\label{df:regular-poles} Let us consider a pole $\bbot\subseteq \overline{\Pi}$. We say that $\bbot$ is \emph{regular} iff for all $(P,e), (Q, f)$ (PWF)-processes, $\{(Q, f)\}^{\perp}\subseteq \{(P, e)\}^{\perp}$ whenever $(P, e)\reduce (Q, f)$. 
\end{definition}
\begin{proposition}\label{pr:PWF-as-CCPA-model} 
  Let us consider a (CPA) $\mathcal{B}:=(\mathbb{B}, \otimes, \|, ()^\perp, {\bf 1}, \preccurlyeq)$ whose pole $\bbot$ is regular. Then $\mathcal{B}$ is a (CCPA) with the definition ${\bf M}(a,b):= \{(a^\uparrow(b).\one, \Delta_\NN)\}^{\perp\perp}$. 
\end{proposition}
\begin{proof}
  By definition, ${\bf M}(a,b)\in\mathcal{S}$ because is not empty. The other combinators ${\bf K}(a), {\bf F}(a,b), {\bf Bl}(a,b), {\bf Br}(a,b), {\bf D}(a,b,c)$ and ${\bf S}(a,b,c)$ are given by {\bf Definition} \ref{df:CCPA} and they are inhabited by their respective encoding on (PWF).  
\end{proof}
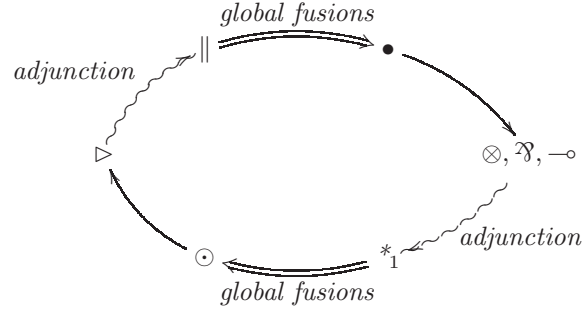
\begin{remark}\label{rk:Interdefinability-of-operations}\ 
  \begin{center}
    $\xymatrix{
      & \| \ar@{=>}@/^/[rr]^{\mbox{global fusions}} &&\bullet\ar@{->}@/^/[dr]& \\
      \rhd \ar@{~>}@/^/[ur]^{\mbox{adjunction}}&&&&\otimes, \parr,\multimap \ar@{~>}@/^/[dl]^{\mbox{adjunction}}\\
      &\odot\ar@{->}@/^/[ul]&& \mbox{*}_1\ar@{=>}@/^/[ll]^{\mbox{global fusions}}&\\
    }$
  \end{center}
  This schemata shows the relation between the logical side (immanent for conjunctive algebras of Miquey) and the concurrent computational side: The relation between both sides is given by the fussion of names.

  \begin{center}
    \begin{tabular}{rcl}
      $A\multimap B$&$:=$&$(A\otimes B^\perp)^\perp$ (internal for (CS))\\
      $A\parr B$&$:=$&$(\mathcal{A}^\perp\mathcal{B}^\perp)^\perp$ (internal for (CS))\\
      $A*_1 B$&$:=$&$\bigcurlywedge\{X\in\mathbb{B}\ |\ A\subseteq B\multimap X\}$ (internal for (CS), building an adjunction $*_1, \multimap$ by completeness)\\
      $A \odot B$&$:=$&$[\iota_2]\big([\iota_1]\mathcal{A} *_1\mathcal{B}\big)$ (defined from $*_1$ by fusions instead of $\iota_1, \iota_2$)\\
      $A\rhd B$&$:=$&$\big(A\odot B^\perp\big)^\perp$ (internal for (CS))\\
      $A\| B$&$:=$&$\bigcurlyvee\{Y\in\mathbb{B}\ |\ A\subseteq B\rhd Y\}$ (internal for (CS), building an adjunction by completeness)\footnote{Here is needed the compatibility between $\|$ and $\bigcurlyvee$}\\
      $A\bullet B$&$:=$&$[\iota_1]\mathcal{A}\|[\iota_2]\mathcal{B}$ (using fusions instead of $\iota_1, \iota_2$)\\
      $A\otimes B$&$:=$&$\big(A\bullet B\big)^{\perp\perp}$ (internal for (CS))\\
    \end{tabular}
  \end{center}
  The right-hand side of the diagram corresponds to the logical connectives $\otimes, \parr, \multimap$ of MLL and the adjoint application of $\multimap$, while $\bullet$ is an auxiliary definition. Operationally, the logical connectives corresponds to parallel composition without communication.
  
  The left-hand side corresponds to the operational semantics of the the realizers. The parallel composition $\|$ corresponds to general parallelism, where communication is allowed. The operation $\lhd$ is the right adjoint of general parallelism while $\odot$ is an auxiliary definition. 
  
  The diagram explains that the link between the logical side and the operational side is given by global fusions.  
\end{remark}
%\section{Annexes}

%\input{annexes.tex} 

\end{document}